\documentstyle[12pt]{article}
\newcommand{\rmd}{{\rm d}} 
\newcommand{\Tr}{\mathop{\rm Tr}} 
\newcommand{\tr}{\mathop{\rm tr}} 
\newcommand{\U}{{\rm U}} 
 
\begin{document}

\begin{flushright}
{UT-02-53 \\
YITP-SB-02-55\\ 
IU-MSTP/53}
\end{flushright}
\vskip 0.5 truecm

\begin{center}
{\Large{\bf Domain wall fermion and CP symmetry breaking}} 
\end{center} 
\vskip .5 truecm 
\centerline{\bf Kazuo Fujikawa} 
\vskip .4 truecm 
\centerline{\it Department of Physics, University of Tokyo} 
\centerline{\it Bunkyo-ku, Tokyo 113, Japan} 
\centerline{\it and} 
\centerline{\it C.N. Yang Institute for Theoretical Physics} 
\centerline{\it State University of New York, Stony Brook, NY 
11794-3840, USA} 
\vskip 0.5 truecm 
\centerline{\bf Hiroshi Suzuki} 
\vskip .4 truecm 
\centerline {\it Department of Mathematical Sciences, 
Ibaraki University} 
\centerline {\it Mito, 310-8512, Japan} 
\makeatletter \@addtoreset{equation}{section} 
\def\theequation{\thesection.\arabic{equation}}
\makeatother

\begin{abstract}
We examine the CP properties of chiral gauge theory defined by 
a formulation of the domain wall fermion, where the light field 
variables $q$ and $\bar q$ together with Pauli-Villars fields 
$Q$ and $\bar Q$ are utilized. It is shown that this domain 
wall representation in the infinite flavor limit $N=\infty$ is 
valid only in the topologically trivial sector, and that the 
conflict among lattice chiral symmetry, strict locality and  CP 
symmetry still persists for finite lattice spacing $a$. The CP 
transformation generally sends one 
representation of lattice chiral gauge theory into another 
representation of lattice chiral gauge theory, resulting in the 
inevitable change of propagators. A modified form of lattice CP 
transformation motivated by the domain wall fermion, which keeps
 the chiral action in terms of the Ginsparg-Wilson fermion 
invariant, is analyzed in detail; this provides an alternative 
way to understand the breaking of CP symmetry at least in the 
topologically trivial sector. We note that the conflict with CP
symmetry could be regarded as a topological obstruction.
We also discuss the issues related
 to the definition of Majorana fermions in connection with the 
supersymmetric Wess-Zumino model on the lattice.   
\end{abstract}

\section{Introduction}

It has been recently shown that CP symmetry in chiral gauge 
theory~\cite{hasenfratz,fis1,fis2} and also the Majorana 
reduction in the presence of chiral symmetric Yukawa 
couplings~\cite{fi1} have a certain conflict with lattice 
chiral symmetry, doubler-free and locality conditions in the 
framework of Ginsparg-Wilson 
operators~\cite{ginsparg}--\cite{fi2}. There exists a closely 
related formulation of lattice fermions which is called the 
domain wall fermion~\cite{kaplan}--\cite{kikukawa3}. In one 
representation of the domain wall fermion in the infinite 
flavor limit, 
the domain wall fermion becomes identical to the overlap 
fermion~\cite{narayanan2,randjbar-daemi,neuberger3}
and thus to the Ginsparg-Wilson fermion. In such a case, the 
conflict with CP symmetry in chiral theory naturally persists if
 one uses the conventional representation of Ginsparg-Wilson 
fermions. 
There are however other representations of the domain wall 
fermion when discussing chiral 
symmetry~\cite{shamir, vranas, kikukawa2}, and 
in those representations (and also in the conventional 
overlap fermion~\cite{narayanan2,randjbar-daemi,neuberger3}), 
the conflict with CP symmetry is less obvious. It is therefore 
desirable to examine in detail how the conflict observed in the 
framework of Ginsparg-Wilson fermions persists in the domain 
wall fermion.

We analyze this issue in a formulation of the domain wall 
fermion where the light field variables $q$ and $\bar q$ 
together with Pauli-Villars fields $Q$ and~$\bar Q$ are 
utilized~\cite{vranas, neuberger2, kikukawa2}. To make this 
analysis as definite as possible, we concentrate on the 
infinite flavor $N=\infty$~limit of the domain wall fermion, 
where chiral symmetry is well-defined. It is shown that this 
representation of the domain wall fermion  is valid only in the 
topologically trivial sector and that the conflict with CP 
symmetry persists. 
We also analyze in detail a modified form of lattice CP 
transformation motivated by the domain wall fermion, which 
keeps the chiral action in the Ginsparg-Wilson fermion 
invariant, and show that CP symmetry is still violated.
In the analysis of CP symmetry, it turns out that topological 
considerations play 
an essential role and, in fact, the conflict with CP symmetry 
could be regarded as a topological obstruction.

In connection with the definition of Majorana fermions and its 
application to supersymmetry, we note a possibility of 
replacing the Pauli-Villars fields in the domain wall 
formulation by the auxiliary field in the Wess-Zumino model. 
In fact this formulation agrees with a past 
suggestion~\cite{fi1, fujikawa2} of the Wess-Zumino action in 
terms of the Ginsparg-Wilson operators.
 
In this paper we take as a basis of our analysis a hermitian 
lattice operator defined by 
\begin{equation}
   H=a\gamma_5D=H^\dagger=aD^\dagger\gamma_5
\end{equation}
where $D$~stands for the lattice Dirac operator and $a$~is the 
lattice spacing. The Ginsparg-Wilson operator is then defined 
by the algebraic relation 
\begin{equation}
   \gamma_5H+H\gamma_5=2H^2
\end{equation}
and its solution agrees with the overlap 
operator~\cite{neuberger} (and its variants).

Although the above simplest form of the Ginsparg-Wilson 
relation is relevant to our analysis of the domain wall fermion,
  the generality of the conflict with CP (or C) symmetry is best 
understood if one considers a more general algebraic 
relation~\cite{fis1}
\begin{equation}
   \gamma_5H+H\gamma_5=2H^2f(H^2)
\end{equation}
where $f(H^2)$ is assumed to be a regular function of~$H^2$ and
$f(H^2)^\dagger=f(H^2)$: $f(x)$~is assumed to be monotonous and 
non-decreasing for~$x\geq0$. The explicit construction of the 
operator~$D$ is known for $f(H^2)=H^{2k}$ with non-negative 
integers~$k$~\cite{fujikawa, fi2}, and $k=0$ gives rise to the 
conventional Ginsparg-Wilson relation~\cite{niedermayer}. In 
our analysis of CP symmetry, the operator defined by 
\begin{equation}
   \Gamma_5=\gamma_5-Hf(H^2)
\end{equation}
or $\gamma_5\Gamma_5$ plays a central role. This operator 
satisfies the relation 
\begin{equation}
   \Gamma_5H+H\Gamma_5=0
\end{equation}
and $\gamma_5\Gamma_5$ vanishes for some momentum variables 
inside the basic Brillouin zone.

This vanishing of $\gamma_5\Gamma_5$ is shown on the general 
ground of locality and species doubler-free conditions of~$H$. 
We here briefly illustrate the basic reasoning, since it is 
closely related to the basic issue of the domain wall fermion: 
One can confirm the relation 
\begin{equation}
   \gamma_5H^2
   =(\gamma_5H+H\gamma_5)H-H(\gamma_5H+H\gamma_5)
+ H^2\gamma_5=H^2\gamma_5
\end{equation} 
which implies $H^2=\gamma_5H^2\gamma_5$ and thus $DH^2=H^2D$. 
The above defining relation~(1.3) is also written as 
\begin{equation}
   \gamma_5H+H\hat\gamma_5=0,\qquad\gamma_5D+D\hat\gamma_5=0
\end{equation}
and $\hat\gamma^2_5=1$ where
\begin{equation}
   \hat\gamma_5=\gamma_5-2Hf(H^2).
\end{equation}
We note that
\begin{equation}
   D\gamma_5\Gamma_5-\gamma_5\Gamma_5D=0
\end{equation}
and also the relation
\begin{eqnarray} \gamma_5\Gamma_5\hat\gamma_5&
=&\gamma_52\Gamma^2_5
-\gamma_5\Gamma_5\gamma_5
=\gamma_5(\gamma_5\Gamma_5+\Gamma_5\gamma_5)
-\gamma_5\Gamma_5\gamma_5
\nonumber\\
&=&\gamma_5(\gamma_5\Gamma_5).
\end{eqnarray}

We now examine the action defined by
\begin{equation}
S=\int\bar\psi D\psi\equiv\sum_{x,y}a^4\bar\psi(x)D(x,y)\psi(y)
\end{equation}
which is invariant under the lattice chiral transformation 
\begin{equation}
   \delta\psi=i\epsilon\hat\gamma_5\psi,\qquad
   \delta\bar\psi=\bar\psi i\epsilon\gamma_5.
\end{equation}
If one considers the field re-definition
\begin{equation}
   q=\gamma_5\Gamma_5\psi,\qquad\bar q=\bar\psi
\end{equation}
the above action is written as
\begin{equation}
   S=\int\bar qD\frac{1}{\gamma_5\Gamma_5}q
\end{equation}
which is invariant under the naive chiral transformation 
\begin{eqnarray}
   \delta q&=&\gamma_5\Gamma_5\delta\psi
   =\gamma_5\Gamma_5i\epsilon\hat\gamma_5\psi
   =i\epsilon\gamma_5q,
\nonumber\\
   \delta\bar q&=&\bar qi\epsilon\gamma_5.
\end{eqnarray}
This chiral symmetry implies the relation
\begin{equation}
   \left\{\gamma_5,D\frac{1}{\gamma_5\Gamma_5}\right\}=0.
\end{equation}
On the basis of the standard argument of the no-go theorem, 
$D/(\gamma_5\Gamma_5)$ and thus $1/(\gamma_5\Gamma_5)$ has 
singularities inside the Brillouin zone for local and species 
doubler-free~$H=a\gamma_5D$. In fact it is 
shown that~\cite{fujikawa}
\begin{equation}
   \Gamma_5=0
\end{equation}
just on top of the would-be species doublers for 
$f(H^2)=H^{2k}$ with non-negative integers~$k$ in the case of 
free fermions and also for the topological modes in the 
presence of instantons. See also Appendix. The field~$q$, which 
plays a central role in the domain-wall 
fermion~\cite{shamir}--\cite{kikukawa2}, is thus ill-defined for these 
configurations.

It is shown that the domain wall variables $q$ and~$\bar q$ in 
the infinite flavor limit satisfy the normal charge conjugation 
properties as well as the continuum chiral symmetry, though 
they are defined in terms of the non-local action. Moreover, 
one can re-write all the correlation functions for $q$ 
and~$\bar q$ in terms of the local variables by using 
$q=\gamma_5\Gamma_5\psi$ and~$\bar q=\bar\psi$~\cite{kikukawa2}.
  One might thus naively expect that we do not encounter any 
difficulty associated with CP and charge conjugation properties.
 The purpose of this paper is to clarify this and related issues.

\section{Domain wall fermions and CP transformation}

\subsection{Chiral properties}

The domain wall fermion is defined by a set of coupled fermion 
fields~\cite{kaplan,shamir} 
\begin{eqnarray}
   a_5{\cal L}_{\rm DW}
   &=&\bar\psi_1
   \left[(\gamma_5a_5H_{\rm W}+1)\psi_1-P_-\psi_2
+\mu P_+\psi_N\right] \nonumber\\
   &&+\sum_{i=2}^{N-1}\bar\psi_i
   \left[(\gamma_5a_5H_{\rm W}+1)\psi_i-P_ -\psi_{i+1}-P_+
\psi_{i-1}\right] \nonumber\\
   &&+\bar\psi_N
   \left[(\gamma_5a_5H_{\rm W}+1)\psi_N+
\mu P_-\psi_1-P_+\psi_{N-1}\right] 
\end{eqnarray} 
where $N$ is chosen to be a positive even integer, and 
\begin{equation}
H_{\rm W}\equiv\gamma_5\left(D_{\rm W}-\frac{m_{0}}{a}\right) 
\end{equation} 
with the Wilson fermion operator $D_{\rm W}$ (with the Wilson 
parameter~$r=1$) and $0<m_0<2$; $a_5$~is the lattice spacing in 
the fifth (or ``flavor'') direction. Note that 
$H_{\rm W}^\dagger=H_{\rm W}$. We use the conventional chiral 
projection operators 
\begin{equation}
   P_\pm=\frac{1\pm\gamma_5}{2}.
\end{equation}
The parameter $\mu$ is chosen to be $\mu=0$ for the domain wall 
variables and $\mu=1$ for the Pauli-Villars variables to 
subtract heavy fermion degrees of freedom. After performing the 
path integral over all the fermion variables one 
obtains~\cite{neuberger2,borici} 
\begin{equation}
   \det\left[\gamma_5(1-a_5H_{\rm W}P_-)\right]^N
   \det\left[(P_--\mu P_+)-T^{-N}(P_+-\mu P_-)\right] 
\end{equation} 
where the transfer operator is given by 
\begin{equation}
   T=\frac{1}{1+a_5H_{\rm W}P_+}(1-a_5H_{\rm W}P_-)
   =\frac{1+{\cal H}_{\rm W}}{1-{\cal H}_{\rm W}} 
\end{equation} 
with 
\begin{equation}
   {\cal H}_{\rm W}=\frac{-1}{2+a_5H_{\rm W}\gamma_5}a_5H_{\rm W}
   =a_5H_{\rm W}\frac{-1}{2+\gamma_5a_5H_{\rm W}}. 
\end{equation} 
Note that both of $H_{\rm W}$ and~${\cal H}_{\rm W}$ are 
hermitian. If one subtracts the contributions of heavy 
fermions (by setting $\mu=1$) from the above determinant 
with~$\mu=0$, one obtains the ``truncated'' overlap or 
Ginsparg-Wilson operator~$D_N$ 
\begin{eqnarray}
   \det aD_N&\equiv&\det\left(P_--T^{-N}P_+\right)
   /\det\left[(P_--P_+)-T^{-N}(P_+-P_-)\right]
\nonumber\\
   &=&\det\left[\frac{1}{2}\left(1+\gamma_5
\frac{1-T^N}{1+T^N}\right)\right]
\end{eqnarray}
where $a$~is the lattice spacing in the four dimensional 
Euclidean space, and the effective Lagrangian for the physical 
fermion 
\begin{equation}
   {\cal L}=\bar\psi D_N\psi.
\end{equation}
Note that $D_N$ is well-defined for $N={\rm even}$ 
and~$a_5/a\ll1$, since $\|a_5H_{\rm W}\|\leq a_5/a$ and 
$T$~is a well-defined hermitian operator.

On the other hand, if one defines the light fermion degrees of 
freedom by~\cite{shamir} 
\begin{equation}
   q\equiv\frac{a}{a_5}(P_-\psi_1+P_+\psi_N),\qquad
   \bar q\equiv\bar\psi_1P_++\bar\psi_NP_-
\end{equation}
and integrates over all the remaining degrees of freedom 
in~eq.~(2.1), one obtains after subtracting the heavy fermion 
contributions by the Pauli-Villars bosonic spinors $Q$ 
and~$\bar Q~$\cite{kikukawa2}, 
\begin{eqnarray}
   {\cal L}_{\rm DW}&=&\bar q\frac{a_5}{a}D^{\rm eff}_Nq
   +\bar Q(1+a_5D^{\rm eff}_N)Q
\nonumber\\
   &=&\bar q\frac{D_N}{1-aD_N}q+\bar Q\frac{1}{1-aD_N}Q. 
\end{eqnarray} 
If one performs the path integral over all the variables, one 
obtains the same result $\det D_N$ as above. We denote this 
Lagrangian (and its $N=\infty$~limit) by~${\cal L}_{\rm DW}$ 
hereafter.

If one takes the infinite flavor limit~$N\to\infty$, one 
obtains (in the limit~$a_5\to0$)
\begin{equation}
   D_N\to D
   =\frac{1}{2a}
   \left(1+\gamma_5\frac{H_{\rm W}}{\sqrt{H^2_{\rm W}}}\right) 
\end{equation} 
which gives the Neuberger's overlap operator~\cite{neuberger} 
satisfying the simplest Ginsparg-Wilson relation 
$\gamma_5D+D\gamma_5=2aD\gamma_5D$. In this limit we can write 
the domain wall fermion as 
\begin{eqnarray}
   {\cal L}_{\rm DW}&=&\bar qD\frac{1}{\gamma_5\Gamma_5}q
   +\bar Q\frac{1}{\gamma_5\Gamma_5}Q
\end{eqnarray}
which is valid for a general class of Ginsparg-Wilson operators.
  In our analysis of CP and related problems, we utilize this 
$N=\infty$ expression.

We here note that
\begin{equation}
1-aD_N=\frac{1}{2}\left(1+\gamma_5\frac{T^N-1}{T^N+1}\right)
\neq0
\end{equation}
since 
\begin{equation}
   \left\|\frac{T^N-1}{T^N+1}\right\|<1
\end{equation}
for finite even $N$ and sufficiently small~$a_5/a$. 
Consequently, $D_N/(1-aD_N)$ is a well-defined and local 
operator (see Ref.\cite{kikukawa4} for the locality of $D_{N}$), 
and 
\begin{equation}
   \left\{\gamma_5,\frac{D_N}{1-aD_N}\right\}\neq0
\end{equation}
since $D_N$ with finite $N$ does not satisfy the Ginsparg-Wilson
  relation. On the other hand, the $N=\infty$~expression 
satisfies 
\begin{equation}
   \left\{\gamma_5,\frac{D}{1-aD}\right\}=0
\end{equation}
and thus the operator $1/(1-aD)$ becomes singular. It is 
interesting that good locality (analytic property) and good 
chiral symmetry for the operator~$D_N/(1-aD_N)$ are traded in 
the limit~$N=\infty$.

The locality of $D_N/(1-aD_N)$ is understood intuitively, since 
the defining Lagrangian of the domain wall fermion for finite 
$N$ couples $N$ fields by the operator $H_{\rm W}$, which causes
  correlation over the finite distances $\sim Na$ in 
4-dimensional Euclidean space. In the limit $N\to\infty$, the 
operator $D/(1-aD)$ could thus become non-local. By subtracting 
the contributions from far apart fields with the 
parameter~$\mu=1$ in the defining Lagrangian, the Pauli-Villars 
fields $Q$ and~$\bar Q$ could restore the
locality: In fact, the singular factor~$1/(1-aD)$ is canceled 
by the Pauli-Villars fields.

The explicit expression of the Lagrangian for chiral gauge 
theory is not specified by the domain wall 
prescription,\footnote{See, however, ref.~\cite{kikukawa3}.} 
since precise chiral symmetry is not defined for finite~$N$. It 
is however natural to analyze chiral theory based on the above 
correspondence to the Ginsparg-Wilson operator 
\begin{eqnarray}
 &&\int{\cal D}\psi{\cal D}\bar\psi\,
\exp\left(\int\bar\psi D\psi\right) \nonumber\\
   &&=\int{\cal D}q{\cal D}\bar q{\cal D}Q{\cal D}\bar Q\,
   \exp\left(\int\bar qD\frac{1}{\gamma_5\Gamma_5}q
   +\int\bar Q\frac{1}{\gamma_5\Gamma_5}Q\right).
\end{eqnarray}

We first note
\begin{equation}
   D=P_+D\hat P_-+P_-D\hat P_+
\end{equation}
with
\begin{equation}
   \hat P_\pm=\frac{1\pm\hat\gamma_5}{2}
\end{equation}
and (by using the relations such as 
$P_\pm\hat P_\pm=P_\pm\gamma_5\Gamma_5$ and~$\gamma_5\Gamma_5 
\hat P_\pm=P_\pm\gamma_5\Gamma_5$) \begin{eqnarray}
   \hat P_-&=&\hat P_-\hat P_-
\nonumber\\
   &=&\hat P_-\frac{1}{\gamma_5\Gamma_5}\gamma_5\Gamma_5
\hat P_- \nonumber\\
   &=&\hat P_-\frac{1}{\gamma_5\Gamma_5}P_-\gamma_5\Gamma_5
\hat P_-
   +\hat P_-\frac{1}{\gamma_5\Gamma_5}P_+\gamma_5\Gamma_5
\hat P_- \nonumber\\
   &=&\hat P_-\frac{1}{\gamma_5\Gamma_5}P_-\hat P_-
\hat P_- \nonumber\\
   &=&\hat P_-\frac{1}{\gamma_5\Gamma_5}P_-P_-\gamma_5\Gamma_5.
\end{eqnarray}

We then have the chiral Lagrangian
\begin{eqnarray}
   {\cal L}_L&=&\int\bar\psi P_+D\hat P_-\psi
\nonumber\\
   &=&\int\bar qP_+D\frac{1}{\gamma_5\Gamma_5}P_-q
\end{eqnarray}
with 
\begin{eqnarray}
   q(x)&=&\gamma_5\Gamma_5\psi(x),
\nonumber\\
   \bar q(x)&=&\bar\psi(x)
\end{eqnarray}
and
\begin{eqnarray}
   \psi_L&\equiv&\hat P_-\psi=\hat P_-
\frac{1}{\gamma_5\Gamma_5}P_-(P_-q)
   =\hat P_-\frac{1}{\gamma_5\Gamma_5}P_-q_L,
\nonumber\\
   \bar\psi_L&\equiv&\bar\psi P_+=\bar qP_+=\bar q_L. 
\end{eqnarray} 
The path integral is then given by taking the Jacobian 
associated with the above change of variables into account, 
\begin{eqnarray}
   &&\int{\cal D}\psi_L{\cal D}\bar\psi_L\,
   \exp\left(\int\bar\psi P_+D\hat P_-\psi\right)
\nonumber\\
   &&=\int{\cal D}q_L{\cal D}\bar q_L{\cal D}Q_L{\cal D}
\bar Q_R\,
   \exp\left(\int\bar qP_+D\frac{1}{\gamma_5\Gamma_5}P_-q
   +\int\bar Q\hat P_-\frac{1}{\gamma_5\Gamma_5}P_-Q\right)
\end{eqnarray}
where we defined the bosonic Pauli-Villars spinors 
\begin{eqnarray}
   &&Q_L(x)=P_-Q(x),
\nonumber\\
   &&\bar Q_R(x)=\bar Q(x)\hat P_-.
\end{eqnarray}
This is consistent if one recalls
\begin{eqnarray}
   &&D\frac{1}{\gamma_5\Gamma_5}
   =P_+D\frac{1}{\gamma_5\Gamma_5}P_-+P_-D
\frac{1}{\gamma_5\Gamma_5}P_+,
\nonumber\\
   &&\frac{1}{\gamma_5\Gamma_5}=\hat P_-
\frac{1}{\gamma_5\Gamma_5}P_-
   +\hat P_+\frac{1}{\gamma_5\Gamma_5}P_+.
\end{eqnarray}

The chiral transformation laws of various fields are defined by 
\begin{eqnarray}
   &&\psi\to e^{i\alpha\hat\gamma_5}\psi,\qquad
   \bar\psi\to\bar\psi e^{i\alpha\gamma_5},
\nonumber\\
   &&q\to e^{i\alpha\gamma_5}q,\qquad\bar q\to
\bar qe^{i\alpha\gamma_5}, \nonumber\\
   &&Q\to e^{i\alpha\gamma_5}Q,\qquad\bar Q\to
\bar Qe^{-i\alpha\hat\gamma_5} 
\end{eqnarray} 
and, similarly, the fermion number transformation by 
\begin{eqnarray}
   &&\psi\to e^{-i\alpha}\psi,\qquad\bar\psi\to\bar\psi 
e^{i\alpha}, \nonumber\\
   &&q\to e^{-i\alpha}q,\qquad\bar q\to\bar qe^{i\alpha}, 
\nonumber\\
   &&Q\to e^{-i\alpha}Q,\qquad\bar Q\to\bar Qe^{i\alpha}. 
\end{eqnarray} 
These transformation rules are fixed if one formally gauges 
those degrees of freedom.

Based on this formulation of chiral gauge theory, we make the 
following observations:

\noindent
(i) One may take the Ginsparg-Wilson variables $\psi$ 
and~$\bar\psi$, which are defined by a local Lagrangian, as the 
primary  variables. One may thus add the source terms to both 
hand-sides of the path integral 
\begin{eqnarray}
   {\cal L}_{\rm source}
&=&\bar\eta_R\psi_{L}+\bar\psi_{L}\eta_R
\nonumber\\   
&=&\bar\eta_R\hat P_-\psi+\bar\psi P_+\eta_R
\nonumber\\
   &=&\bar\eta_R\hat P_-\frac{1}{\gamma_5\Gamma_5}P_-(P_-q)+
\bar qP_+\eta_R. 
\end{eqnarray} 
To avoid the singularities appearing in various expressions for 
domain wall variables, one needs to work in the functional space
{\em without} the modes 
\begin{equation}
   \Gamma_5\varphi_n=0
\end{equation}
in the domain wall representation. We here recall that the 
index for the Ginsparg-Wilson operator is given 
by~\cite{hasenfratz3,luscher,ky,chiu,fujikawa3}
\begin{equation}
   \Tr\Gamma_5=n_+-n_-=N_--N_+
\end{equation}
where $n_\pm$ stand for the modes 
$H\varphi_n=a\gamma_5D\varphi_n=0$
with~$\gamma_5\varphi_n=\pm\varphi_n$, respectively, and 
$N_\pm$ stand for the modes 
$\Gamma_5\varphi_n=[\gamma_5-Hf(H^2)]\varphi_n=0$
with~$\gamma_5\varphi_n=\pm\varphi_n$, respectively. 
See~Appendix. The constraint $N_+=N_-=0$ thus implies that we 
work in the {\em topologically trivial sector} 
with~$\Tr\Gamma_5=0$. 
This constraint is consistent with the above fermion number 
transformation: For the Ginsparg-Wilson variables, we obtain 
the Jacobian factor\footnote{The path integral for chiral 
non-Abelian gauge theory has not been completely understood yet.
  But the chiral $\U(1)$ anomaly and associated index are 
insensitive to the details of the path integral measure.} 
\begin{eqnarray}
   \ln J_\psi=i\alpha\Tr\hat P_--i\alpha\Tr P_+ 
=-i\alpha\Tr\Gamma_5 
\end{eqnarray} 
whereas for the domain wall variables $q$ and~$\bar q$, we 
obtain 
\begin{eqnarray}
   \ln J_q=i\alpha\Tr P_--i\alpha\Tr P_+=-i\alpha\Tr\gamma_5. 
\end{eqnarray} 
We thus have $\Tr\Gamma_5=\Tr\gamma_5=0$.

\noindent
(ii) One may take the domain wall variables as the primary 
variables and add the source terms 
\begin{eqnarray}
   {\cal L}_{\rm source}
&=&\bar\eta_R q_{L}+\bar q_{L}\eta_R \nonumber\\
&=&\bar\eta_RP_-q+\bar qP_+
\eta_R \nonumber\\
   &=&\bar\eta_RP_-\gamma_5\Gamma_5\psi+\bar\psi P_+\eta_R
\nonumber\\
&=&\bar\eta_RP_-\psi_{L}+\bar\psi_{L}\eta_R 
\end{eqnarray} 
where we used $P_{-}\gamma_{5}\Gamma_{5}=P_{-}\hat{P}_{-}$.
One might attempt to interpret the chiral domain wall 
representation with these source terms in the following way: In 
any fermion loop diagram such as in the determinant factor, we 
combine the variables $q$,~$\bar q$ and $Q$,~$\bar Q$ together 
and obtain the determinant without the 
$1/(\gamma_{5}\Gamma_{5})$ factor. For the external
  fermion lines connected to the source terms, we use the 
variables $q_{L}$ and~$\bar q_{L}$ by replacing those variables 
later by
 $P_{-}\psi_{L}$ and~$\bar\psi_{L}$. By this way, we do not 
encounter any singularity even in topologically non-trivial 
sectors. This is perhaps the simplest view based on the domain 
wall fermion. This view however has a fatal difficulty in 
topological properties, namely, one cannot generate the fermion 
number anomaly by a transformation of variables $q_{L}$ 
and~$\bar q_{L}$ alone 
\begin{eqnarray}
 \ln J_q=i\alpha\Tr P_--i\alpha\Tr P_+=-i\alpha\Tr\gamma_5=0 
\end{eqnarray} 
if one works in the complete functional space in topologically
non-trivial sectors.
We thus have to exclude the modes $\Gamma_5\varphi_n=0$ by hand,
  for example, but we may still work in  all the topological 
sectors on the basis of the domain wall variables $q_{L}$ 
and~$\bar q_{L}$. In this case, the topological properties are 
maintained since one can confirm~\cite{fujikawa3} 
\begin{equation}
   \Tr\nolimits'\gamma_5=n_+ -n_-
\end{equation}
where $\Tr\nolimits'$ is taken in the functional space with the 
modes~$\Gamma_5\varphi_n=0$ excluded. This index relation has 
the same form as in continuum theory~\cite{fujikawa4}. This 
exclusion of the modes~$\Gamma_5\varphi_n=0$ is consistent 
with the
replacement~$q\to\gamma_5\Gamma_5\psi$ since the 
factor~$\gamma_5\Gamma_5$ projects out those 
modes~\cite{narayanan, chandrasekharan}. 
This operation is however apparently 
non-local.\footnote{The exclusion of the modes 
$\Gamma_5\varphi_n(x)=0$ in all the topological sectors is 
apparently a non-local operation in 
spacetime, though it is a local operation in the ``mode space'',
  since the functional value of~$\varphi_n(x)$ is fixed over 
the entire space once its value at one point is fixed. The 
exclusion of the modes $\Gamma_5\varphi_n(x)=0$ by hand 
corresponds to the exclusion of would-be species doublers by 
hand.}

Based on these considerations we conclude that the domain wall 
fermion representation in the limit $N=\infty$, where chiral 
symmetry is well defined, is valid as a local field theory (in 
the above interpretation) only in the topologically trivial 
sector with $\Tr\Gamma_5=0$. The primary variables, which 
describe the full physical contents expressed by various 
correlation functions in topologically trivial as well as 
non-trivial sectors, are thus given by the Ginsparg-Wilson 
fermions $\psi_{L}$ and~$\bar\psi_{L}$, and hereafter we 
analyze the domain wall representation in the topologically 
trivial sector with the source terms (2.29)~added. Note that 
source terms specify the correlation functions, and
the component $P_{+}\psi_{L}$ is missing in (2.34); to maintain
the consistency of internal and external fermion lines, which is 
related to unitarity, we need to use the source (2.29).

In passing, another interesting representation, which is 
equivalent to the domain wall fermion, is given by 
\begin{eqnarray}
   &&\int{\cal D}\psi_L{\cal D}\bar\psi_L\,
   \exp\left(\int\bar\psi P_+D\hat P_-\psi\right)
\nonumber\\
  &&=\int{\cal D}q_L{\cal D}\bar q_L{\cal D}S_L{\cal D}\bar S_R\,
   \exp\left(\int\bar qP_+D\frac{1}{\gamma_5\Gamma_5}P_-q
   +\int\bar SP_-\gamma_5\Gamma_5\hat P_-S\right) 
\end{eqnarray} 
where we defined the {\it fermionic\/} auxiliary fields 
\begin{eqnarray}
   &&S_L(x)=\hat P_-S(x),
\nonumber\\
   &&\bar S_R(x)=\bar S(x)P_-
\end{eqnarray}
by noting
\begin{equation}
   P_+D\hat P_-=\left(P_+D\frac{1}{\gamma_5\Gamma_5}P_-\right)
   \left(P_-\gamma_5\Gamma_5\hat P_-\right)
\end{equation}
and
\begin{equation}
   \gamma_5\Gamma_5
   =P_-\gamma_5\Gamma_5\hat P_-+P_+\gamma_5\Gamma_5\hat P_+. 
\end{equation} 
The vector-like theory is then defined by 
\begin{eqnarray}
   &&\int{\cal D}\psi{\cal D}\bar\psi\,\exp\left(\int\bar\psi 
D\psi\right) \nonumber\\
   &&=\int{\cal D}q{\cal D}\bar q{\cal D}S{\cal D}\bar S\,
   \exp\left(\int\bar qD\frac{1}{\gamma_5\Gamma_5}q
   +\int\bar S\gamma_5\Gamma_5S\right).
\end{eqnarray}
This representation is applicable only to the topologically 
trivial sector, but it turns out to be convenient when we 
discuss Majorana fermions in the domain wall representation 
later.

\subsection{CP symmetry in chiral gauge theory}

We recall the charge conjugation properties of various 
operators. We employ the convention of the charge conjugation 
matrix~$C$ 
\begin{eqnarray}
   &&C\gamma^\mu C^{-1}=-(\gamma^\mu)^T,
\\ 
   &&C\gamma_5C^{-1}=\gamma^T_5,
\\ 
   &&C^\dagger C=1,\qquad C^T=-C.
\end{eqnarray}
We then have\footnote{We define the CP operation by 
$W=C\gamma_0=\gamma_2$ with hermitian $\gamma_2$ and the CP 
transformed gauge field by $U^{\rm CP}$, and then 
$WD(U^{\rm CP})W^{-1}=D(U)^T$. If the parity is realized in the 
standard way, we have $CD(U^{\rm C})C^{-1}=D(U)^T$.} 
\begin{eqnarray}
   &&WD(U^{\rm CP})W^{-1}=D(U)^T,\qquad
   W\gamma_5\Gamma_5(U^{\rm CP})W^{-1}=[\gamma_5\Gamma_5(U)]^T,
\nonumber\\
   &&WH(U^{\rm CP})W^{-1}=-[\gamma_5H(U)\gamma_5]^T,\qquad
   WH^2(U^{\rm CP})W^{-1}=[H^2(U)]^T,
\nonumber\\
 &&W\Gamma_5(U^{\rm CP})W^{-1}=-[\gamma_5\Gamma_5(U)\gamma_5]^T,
\nonumber\\
   &&W(\Gamma_5/\Gamma)(U^{\rm CP})W^{-1}
   =-[(\gamma_5\Gamma_5\gamma_5/\Gamma)(U)]^T
\end{eqnarray}
where 
\begin{equation}
   \Gamma=\sqrt{\Gamma^2_5}=\sqrt{(\gamma_5\Gamma_5\gamma_5)^2}
   =\sqrt{1-H^2f^2(H^2)}.
\end{equation}
Here we imposed the relation $WD(U^{\rm CP})W^{-1}=D(U)^T$ 
or~$[CD(U)]^T=-CD(U^{\rm C})$ which is consistent with the 
defining Ginsparg-Wilson relation.

We also have the properties
\begin{eqnarray}
   &&W\hat\gamma_5(U^{\rm CP})W^{-1}
   =-\left[\gamma_5\hat\gamma_5(U)\gamma_5\right]^T
\nonumber\\
   &&W\frac{1}{\gamma_5\Gamma_5(U^{\rm CP})}W^{-1}
   =\left[\frac{1}{\gamma_5\Gamma_5(U)}\right]^T.
\end{eqnarray}
We now examine the CP symmetry in chiral gauge theory 
\begin{equation}
   {\cal L}_L=\bar\psi_LD\psi_L
\end{equation}
where we defined the (general) projection operators 
\begin{eqnarray}
   &&D=\bar P_LDP_L+\bar P_RDP_R,
\nonumber\\
 &&\psi_{L,R}=P_{L,R}\psi,\qquad\bar\psi_{L,R}
=\bar\psi\bar P_{L,R}. 
\end{eqnarray} 
Under the standard CP transformation\footnote{The vector-like 
theory is invariant under this CP transformation.} 
\begin{eqnarray}
   &&\bar\psi\to\psi^TW,
\nonumber\\
   &&\psi\to-W^{-1}\bar\psi^T
\end{eqnarray}
the chiral action is invariant only if
\begin{eqnarray}
   &&WP_LW^{-1}=\bar P^T_L,\qquad W\bar P_LW^{-1}=P^T_L. 
\end{eqnarray} 
It was shown elsewhere that the unique solution for this 
condition in the framework of the Ginsparg-Wilson operators is 
given by~\cite{fis1} 
\begin{eqnarray}
   P_{L,R}=\frac{1}{2}(1\mp\Gamma_5/\Gamma),
\nonumber\\
   \bar P_{L,R}=\frac{1}{2}(1\pm\gamma_5\Gamma_5\gamma_5/\Gamma),
\end{eqnarray}
but these projection operators suffer from singularities 
in~$1/\Gamma$. Namely, it is impossible to maintain the 
manifest invariance of the local and chiral Lagrangian under the
  CP transformation~\cite{hasenfratz,fis1}.\footnote{This
however shows that one can maintain manifest CP invariance, if 
one ignores the singularities associated 
with~$\gamma_5\Gamma_5=0$.}

If one stays in the well-defined local Lagrangian 
\begin{equation}
   \int{\cal L}_L=\int\bar\psi P_+D\hat P_-\psi
\end{equation}
it is not invariant under the standard CP transformation as 
\begin{eqnarray}
   &&WP_\pm W^{-1}=P^T_\mp\neq\hat P^T_\mp(U),
\nonumber\\
   &&W\hat P_\pm(U^{\rm CP})W^{-1}
   =\frac{1\mp\left[\gamma_5\hat\gamma_5(U)\gamma_5\right]^T}{2}
   =\left[\gamma_5\hat P_\mp(U)\gamma_5\right]^T\neq P^T_\mp, 
\nonumber\\
   &&\left[WP_+D(U^{\rm CP})\hat P_-(U^{\rm CP})W^{-1}\right]^T
   =\gamma_5\hat P_+(U)\gamma_5D(U)P_-
\nonumber\\
   &&=P_+D(U)\hat P_-(U)-D(U)[\gamma_5-\Gamma_5(U)]
\neq P_+D\hat P_-. 
\end{eqnarray} 
Since one can show that 
\begin{eqnarray}
   &&\left(\gamma_5\hat P_\pm\gamma_5\right)
   \left(\gamma_5\hat P_\pm\gamma_5\right)
   =\left(\gamma_5\hat P_\pm\gamma_5\right),
\nonumber\\
   &&D=\left(\gamma_5\hat P_+\gamma_5\right)DP_-
   +\left(\gamma_5\hat P_-\gamma_5\right)DP_+,
\end{eqnarray}
the CP transformation actually maps one specific representation 
of chiral gauge theory to another representation of chiral 
gauge theory 
\begin{equation}
   \int{\cal L}=\int\bar\psi P_+D(U)\hat P_-(U)\psi\to
   \int{\cal L}=\int\bar\psi\gamma_5\hat P_+(U) \gamma_5D(U)
P_-\psi 
\end{equation} 
based on the {\it same\/} vector-like theory defined by the 
lattice operator~$D$.

It may be appropriate to recall here the essence of our previous
  analysis~\cite{fis2}. The functional space in our problem is 
naturally spanned by the eigenfunctions of the basic hermitian 
operator $H=a\gamma_5D$ 
\begin{equation}
   H\varphi_n=\lambda_n\varphi_n.
\end{equation}
However this eigenvalue equation is gauge covariant as are all 
the quantities in the gauge invariant lattice regularization. 
To accommodate the gauge non-covariant quantities such as a 
consistent form of anomaly, one defines the path integral in a 
specific topological sector specified by~$M$ by 
\begin{equation}
   Z_M(U)=\exp\left[
   i\vartheta_M\left(
   \langle w_n|v_m\rangle;\langle\bar w_n|\bar v_m\rangle\right)
 \right]
   \int\prod_{n,l}\rmd a_n\rmd\bar a_l\,
   \exp\left[\int{\cal L}(\bar\psi,\psi,U)\right] 
\end{equation} 
where we expanded fermionic variables as 
\begin{eqnarray}
   &&\hat P_-\psi=\sum_na_nw_n,
\nonumber\\
   &&\bar\psi P_+=\sum_n\bar a_n\bar w_n.
\end{eqnarray}
The basis vectors $\{w_n\}$ and $\{\bar w_n\}$, which satisfy 
\begin{equation}
   \hat P_-w_n=w_n,\qquad\bar w_nP_+=\bar w_n,
\end{equation}
are suitable linear combinations of $\{\varphi_n\}$ 
and~$\{\varphi_n^\dagger\}$, respectively. 
The ``measure factor''~$\vartheta_M$, which is the Jacobian for 
the transformation from {\it ideal\/} bases $\{v_n\}$ 
and~$\{\bar v_n\}$ to the bases specified by~$H$ and thus 
crucially depends on the ideal bases,\footnote{The measure 
factor is thus chosen to be a constant for the ideal bases.} is 
not specified at this stage and it is later determined by 
imposing several physical conditions.

When one considers the change of fermionic variables which 
formally corresponds to gauge transformation, $\psi\to\psi'$ 
and~$\bar\psi\to\bar\psi'$, the expansion coefficients with the 
fixed basis vectors are transformed as $\{a_n\}\to\{a_n'\}$ 
and~$\{\bar a_n\}\to\{\bar a_n'\}$. Since the naming of path 
integral variables does not matter, one obtains the identity 
\begin{eqnarray}
&&\exp\left[i\vartheta_M\left(
\langle w_n|v_m\rangle;\langle\bar w_n|\bar v_m \rangle\right)
\right]
   \int\prod_{n,l}\rmd a_n\rmd\bar a_l
   \exp\left[\int{\cal L}(\bar\psi,\psi,U)\right]
\nonumber\\
&&=\exp\left[i\vartheta_M\left(
\langle w_n|v_m\rangle;\langle\bar w_n|\bar v_m\rangle\right) 
\right]
   \int\prod_{n,l}\rmd a_n'\rmd\bar a_l'
   \exp\left[\int{\cal L}(\bar\psi',\psi',U)\right]. 
\end{eqnarray} 
In this form of identity, the Jacobian of path integral measure 
gives a lattice version of covariant anomaly and the variation 
of the action gives the divergence of covariant current. The 
gauge covariant fermion number anomaly is naturally derived in 
this way.

If one performs the simultaneous gauge transformation of the 
link variables~$U$ in the above path integral~$Z_{M}(U)$, the 
action 
becomes invariant but one needs to take into account the 
variation of the
measure factor~$\delta\vartheta_M(\langle w_n|v_m\rangle; 
\langle\bar{w}_n|\bar{v}_m\rangle)$ induced by the gauge 
transformation of~$U$. This variation~$\delta\vartheta_M$ 
converts the covariant anomaly to a lattice form of consistent 
anomaly, which is one of the requirements on the measure factor. 
In the anomaly-free theory, $\delta\vartheta_M$ should 
completely cancel the non-vanishing Jacobian arising from 
lattice artifacts. The current associated 
to~$\delta\vartheta_M$ should 
be local and satisfy several other requirements: The existence 
proof of such a measure factor~$\vartheta_M$ amounts to a 
definition of lattice chiral gauge 
theory~\cite{luscher2}--\cite{luscher4}.

A characteristic property of the Ginsparg-Wilson algebra is 
that among the eigenfunctions of 
$H\varphi_n=\lambda_n\varphi_n$ the eigenstates corresponding to
  zero modes and also those eigenstates corresponding to the 
largest values of~$|\lambda_n|$ are chosen to be the 
simultaneous eigenstates of~$\gamma_5$, and that those states 
corresponding to the largest values of~$|\lambda_n|$ are 
annihilated by~$\Gamma_5\varphi_n=0$. By noting 
\begin{equation}
   \hat P_\pm=\frac{1\pm\hat\gamma_5}{2}=P_\mp\pm\Gamma_5
\end{equation}
$\hat P_\pm$ is replaced by $P_\mp$ when acting on the modes 
annihilated by~$\Gamma_5$. We also have the chirality sum rule 
arising from~$\Tr\gamma_5=0$, $n_++N_+=n_-+N_-$, where $n_\pm$ 
and~$N_\pm$ stand for the numbers of zero modes and largest 
eigenmodes with chirality~$\pm$, respectively. See~Appendix.

In terms of the eigenfunctions of the basic operator~$H$, one 
can describe the change of the action under the standard CP 
transformation as follows: One starts with 
\begin{equation}
   \int{\cal L}=\int\bar\psi P_+D(U)\hat P_-(U)\psi 
\end{equation} 
which is characterized by 
\begin{equation}
   \bar\psi=\pmatrix{n_+\cr N_+\cr},\qquad
   \psi=\pmatrix{n_-\cr N_+\cr},\qquad
   n_+-n_-=\Tr\Gamma_5(U).
\end{equation}
Here we write only the number of simultaneous chiral eigenstates
  explicitly, since the same number of eigenstates belonging to 
other eigenvalues are included in~$\psi$ and~$\bar\psi$. The CP 
conjugate theory is defined by 
\begin{equation}
   \int{\cal L}^{\rm CP}=\int\bar\psi^{\rm CP}P_+D(U^{\rm CP})
   \hat P_-(U^{\rm CP})\psi^{\rm CP}
\end{equation}
which is characterized by 
\begin{equation}
\bar\psi^{\rm CP}=\pmatrix{n_+^{\rm CP}\cr N_+^{\rm CP}\cr}, 
\qquad
   \psi^{\rm CP}=\pmatrix{n_-^{\rm CP}\cr N_+^{\rm CP}\cr}, 
\qquad
   n_+^{\rm CP}-n_-^{\rm CP}=\Tr\Gamma_5(U^{\rm CP}) 
=-\Tr\Gamma_5(U). 
\end{equation} 
A regular re-naming of fermionic variables 
in~${\cal L}^{\rm CP}$ 
\begin{equation}
   \bar\psi^{\rm CP}=\psi'W,
\qquad\psi^{\rm CP}=-W^{-1}\bar\psi' 
\end{equation} 
gives rise to 
\begin{equation}
   \int{\cal L}^{\rm CP}
   =\int\bar\psi'\gamma_5\hat P_+(U)\gamma_5D(U)P_-\psi' 
\end{equation} 
which is characterized by 
\begin{equation}
   \bar\psi'=\pmatrix{n_+'=n_-^{\rm CP}\cr N_-'=N_+^{\rm CP}}, 
\qquad
   \psi'=\pmatrix{n_-'=n_+^{\rm CP}\cr N_-'=N_+^{\rm CP}\cr}, 
\qquad
   n_+'-n_-'=\Tr\Gamma_5(U).
\end{equation}
This analysis suggests that we may define the CP transformed 
theory by means of the chiral theory defined by projection 
operators $P_-$ and~$\gamma_5\hat P_+\gamma_5$. It is shown that
  this is in fact consistent including the measure 
factor~\cite{fis2}.

When one compares the original theory to the CP transformed 
theory, the topological index is identical for these two 
theories. Although the number of heaviest modes is different 
$N_-'=N_-\neq N_+$ in general, one may expect that these two 
theories when summed over all the topological sectors give rise 
to an identical result. After all, it should not matter how one 
chooses a specific chiral projection of the original vector-like
  theory specified by~$D$, as long as it is not singular. (The 
continuum limit is expected to be identical, if it is 
well-defined.) This expectation is in fact born out by a 
detailed analysis, and the difference $N_-'\neq N_+$ is taken 
care of by suitably choosing the weight factors for different 
topological sectors when summing those sectors~\cite{fis2}. 
The different actions however give rise to different propagators
  (for finite $a$) 
\begin{equation}
   \left\langle\psi_{L}(x)\bar\psi_{L}(y)\right\rangle
=\hat P_-\frac{1}{D}P_+
   \to P_-\frac{1}{D}\gamma_5\hat P_+\gamma_5
\neq\hat P_-\frac{1}{D}P_+ 
\end{equation} 
which manifest CP breaking in this formulation. From this view 
point, if one chooses projection operators for which the chiral 
theories before and after CP transformation coincide, one 
inevitably encounters a singularity in the topologically 
non-trivial sector because of $N_-'\neq N_+$. The conflict with 
CP symmetry could thus be regarded as a topological obstruction.

The CP non-invariance in the action level persists in the domain
  wall representation 
\begin{equation}
   \int{\cal L}_L=
   \int\bar qP_+D\frac{1}{\gamma_5\Gamma_5}P_-q
   +\int\bar Q\hat P_-\frac{1}{\gamma_5\Gamma_5}P_-Q.
\end{equation}
A natural definition of CP transformation is 
\begin{eqnarray}
   &&\bar q\to q^TW,
\nonumber\\ 
   &&q\to-W^{-1}\bar q^T,
\nonumber\\
   &&\bar Q\to Q^TW,
\nonumber\\ 
   &&Q\to-W^{-1}\bar Q^T.
\end{eqnarray}
This transformation leaves the vector-like theory invariant up 
to the overall signature of the second term in~eq.~(2.17), which
 is 
immaterial.\footnote{One may define a transformation law 
$Q\to W^{-1}\bar Q^T$, for example, to keep the 
action invariant. In this case, however, the CP transformation 
applied twice gives rise to $Q\to-Q$ and~$\bar Q\to-\bar Q$.} 
We note that one cannot keep $Q$ and~$\bar Q$ invariant under 
CP since the gauge field is transformed under CP by 
\begin{equation}
   W\frac{1}{\gamma_5\Gamma_5(U^{\rm CP})}W^{-1}
   =\left[\frac{1}{\gamma_5\Gamma_5(U)}\right]^T.
\end{equation}
The part containing the field $q$ and~$\bar q$ in the above 
chiral Lagrangian is invariant under the CP transformation, but 
the part containing $Q$ and~$\bar Q$ is not invariant under the 
CP transformation 
\begin{eqnarray}
   \left(W\hat P_-\frac{1}{\gamma_5\Gamma_5}P_-W^{-1}\right)^T
   =P_+\frac{1}{\gamma_5\Gamma_5}\gamma_5\hat P_+\gamma_5
   =\hat P_-\frac{1}{\gamma_5\Gamma_5}P_-+\gamma_5
   \neq\hat P_-\frac{1}{\gamma_5\Gamma_5}P_-.
\end{eqnarray}
The conflict with CP symmetry persists as far as the invariance
  of the action is concerned. We recall that $Q$ and~$\bar Q$ 
are  essential to maintain the full physical contents described 
by the variables $\psi$ and~$\bar\psi$. To analyze the effects of
CP violation in the sector of $Q$ and $\bar{Q}$ precisely, we 
discuss a modified CP transformation for $\psi_{L}$ and 
$\bar{\psi}_{L}$ in the next section, which includes the effects
of both of $q,\ \bar{q}$ and $Q,\ \bar{Q}$.
 
As we noted above, this CP transformation is regarded as a 
change of representation specified by 
$P_\pm$ and~$\hat P_\pm$ to another representation specified by 
$\gamma_5\hat P_\pm\gamma_5$ and~$P_\pm$. To be specific, we 
have 
\begin{eqnarray}
   &&\gamma_5\hat P_+\gamma_5
   =(\gamma_5\Gamma_5)P_+P_+
   \left(\frac{1}{\gamma_5\Gamma_5}\right)\gamma_5
\hat P_+\gamma_5, \nonumber\\
   &&\bar\psi\gamma_5\hat P_+\gamma_5
   =\bar\psi(\gamma_5\Gamma_5)P_+P_
+\left(\frac{1}{\gamma_5\Gamma_5}\right)
   \gamma_5\hat P_+\gamma_5
   =\bar qP_+P_+\left(\frac{1}{\gamma_5\Gamma_5}\right)
   \gamma_5\hat P_+\gamma_5,
\nonumber\\
   &&{\cal L}=\bar\psi\gamma_5\hat P_+\gamma_5DP_-\psi
   =\bar qP_+\left(\frac{1}{\gamma_5\Gamma_5}\right)DP_-q
\end{eqnarray}
where we defined
\begin{equation}
   \bar q=\bar\psi\gamma_5\Gamma_5,\qquad q=\psi. 
\end{equation} 
We thus have 
\begin{eqnarray}
   &&\int{\cal D}\psi_L{\cal D}\bar\psi_L\,
   \exp\left(\int\bar\psi\gamma_5\hat P_+\gamma_5DP_-\psi\right)
  \nonumber\\ 
   &&=\int {\cal D}q_L{\cal D}\bar q_L{\cal D}Q_R{\cal D}
\bar Q_L\,
   \exp\left[\int\bar qP_+\left(\frac{1}{\gamma_5\Gamma_5}
\right)DP_-q
   +\int\bar QP_+\frac{1}{\gamma_5\Gamma_5}\gamma_5
\hat P_+\gamma_5Q\right] 
\end{eqnarray} 
where the action for $q$ and~$\bar q$ formally retains the form 
before the CP transformation, though the definitions of $q$ 
and~$\bar q$ in terms of $\psi$ and~$\bar\psi$ are not the same 
as before. To the extent one can define the left-hand side 
consistently, one can define the right-hand side consistently 
except for topological properties. The source terms and the 
resulting propagator, which need to be defined in terms of the 
local variables $\psi$ and~$\bar\psi$, however change under the 
CP transformation as in (2.70). We re-iterate that 
the variables $q$ and~$\bar q$ cannot describe the essential 
properties such as the fermion number non-conservation and 
chirality selection rules (in vector-like theory), which are 
described by the local variables $\psi$ and~$\bar\psi$.

The same conclusion is obtained, namely, CP non-invariance of 
the action, for the ``fermionic'' representation of the domain 
wall fermion 
\begin{equation}
   {\cal L}_L=\int\bar qP_+D\frac{1}{\gamma_5\Gamma_5}P_-q
   +\int\bar SP_-\gamma_5\Gamma_5\hat P_-S
\end{equation}
if one assigns the natural CP transformation law 
\begin{eqnarray}
   &&\bar S\to S^TW,
\nonumber\\ 
   &&S\to-W^{-1}\bar S^T
\end{eqnarray}
which keeps the vector-like theory invariant. We then have 
\begin{equation}
   \left(WP_-\gamma_5\Gamma_5\hat P_-W^{-1}\right)^T
   =\gamma_5\hat P_+\gamma_5\gamma_5\Gamma_5P_+
   =P_-\gamma_5\Gamma_5\hat P_-+\gamma_5\Gamma_5^2
   \neq P_-\gamma_5\Gamma_5\hat P_-
\end{equation}
which is again interpreted as a change of representation of 
lattice chiral theory based on the same vector-like theory.

\section{Modified lattice CP for Ginsparg-Wilson operators}

The part of the Lagrangian for chiral domain wall fermions in 
(2.71), 
which includes the light variables $q$ and~$\bar q$, is 
invariant under CP transformation. This property together with 
$q=\gamma_5\Gamma_5\psi$ and~$\bar q=\bar\psi$ suggest a 
modified lattice CP transformation (which is fixed by first 
going to~$q$ and then coming back to~$\psi$ after CP operation)
\begin{eqnarray}
   \psi_L&=&\hat P_-\psi\to\psi_L^{\rm CP}  
=-W^{-1}\left[\bar\psi_L\frac{1}{\gamma_5\Gamma_5(U)}\right]^T,
\nonumber\\
   \bar\psi_L&=&\bar\psi P_+\to\bar\psi_L^{\rm CP}
   =[\gamma_5\Gamma_5(U)\psi_L]^TW
\end{eqnarray}
for the chiral theory defined in terms of the Ginsparg-Wilson 
fermion 
\begin{equation}
   {\cal L}_L=\bar\psi_LD\psi_L
   =\bar\psi\frac{1+\gamma_5}{2}D\frac{1-\hat\gamma_5}{2}\psi.
\end{equation}
One can confirm that the chiral Lagrangian is invariant 
under the above modified lattice CP transformation. If one 
establishes that the Jacobian for the above modified CP 
transformation 
gives unity, all the effects of CP violation (or abnormal C) 
effects appear in the propagator, which is derived by 
considering source terms (2.29) for $\psi_L$ and~$\bar\psi_L$, 
\begin{eqnarray}
\int{\cal L}^{CP}_{source}&=&
\int[\bar{\eta}^{CP}_{R}\psi^{CP}_{L}+\bar{\psi}^{CP}_{L}
\eta^{CP}_{R}]
\nonumber\\
&=&\int[\bar{\eta}_{R}\gamma_{5}\Gamma_{5}(U)\psi_{L}
+\bar{\psi}_{L}\frac{1}{\gamma_{5}\Gamma_{5}(U)}\eta_{R}],
\end{eqnarray}
and the propagator becomes after CP transformation
\begin{equation}
(\gamma_5\Gamma_5)\hat P_-\frac{1}{D}P_
+\frac{1}{(\gamma_5\Gamma_5)}
   =P_-\frac{1}{D}\gamma_5\hat P_+\gamma_5
   \neq\hat P_-\frac{1}{D}P_+
\end{equation}
in pure chiral gauge theory, to be consistent with our previous 
result~\cite{fis2}. We here assumed the natural CP transformation
for the source functions, 
$\bar{\eta}_{R}\rightarrow \eta^{T}_{R}W$ and 
$\eta_{R}\rightarrow -W^{-1}\bar{\eta}^{T}_{R}$.

This analysis turned out to be rather limited in its scope and 
it is applicable only to the topologically trivial sector, as 
it is directly related to the domain wall representation. It is,
  however, nice to examine the modified lattice CP 
transformation, since, after all, the invention of a lattice 
version of chiral transformation was the starting point of the 
analysis of lattice chiral gauge theory. Also, this analysis 
illustrates an alternative picture about what is going on in 
the analysis of CP symmetry, together with general topological 
complications associated with the transformation which keeps 
action invariant.

In passing, we note that the vector-like theory defined by the 
Ginsparg-Wilson fermion is invariant under the modified CP 
transformation 
\begin{equation}
   \psi\to-\frac{1}{\gamma_5\Gamma_5}W^{-1}\bar\psi^T=-W^{-1}
   \left(\bar\psi\frac{1}{\gamma_5\Gamma_5}\right)^T,\qquad
\bar\psi\to\psi^T(\gamma_5\Gamma_5)^TW=(\gamma_5\Gamma_5\psi)^TW
\end{equation}
and the Jacobian for this transformation is unity up to the 
possible singularity associated with $1/(\gamma_5\Gamma_5)$.

We now present a precise analysis of the Jacobian factor 
associated with the above modified CP transformation in chiral 
gauge theory, and show that we arrive at precisely the same 
conclusion, at least in the topologically trivial sector, as in 
our previous analysis based on the more conventional CP 
transformation~\cite{fis2}. The analysis in this section also 
provides some of the mathematical details briefly sketched in 
the previous section.

We start with the definition of expectation values in the 
fermion sector of the chiral gauge theory 
\begin{equation}
   \left\langle{\cal O}\right\rangle
   =\int{\cal D}\psi_L{\cal D}\bar\psi_L\,{\cal O}\,
   \exp\left(\int\bar\psi_LD\psi_L\right)
\end{equation}
and
\begin{equation}
   {\cal D}\psi_L{\cal D}\bar\psi_L
=\prod_j\rmd c_j\prod_k\rmd\bar c_k. 
\end{equation} 
In this expression, $c_j$ and~$\bar c_k$ are the expansion 
coefficients of fermion fields: 
\begin{equation}
   \psi_L(x)=\sum_jv_j(x)c_j,\qquad
   c_j=(v_j,\psi_L)\equiv a^4\sum_x v_j^\dagger(x)\psi_L(x) 
\end{equation} 
and 
\begin{equation}
   \bar\psi_L(x)=\sum_k\bar c_k\bar v_k(x),\qquad
   \bar c_k=(\bar\psi_L^\dagger,\bar v_k^\dagger)
   \equiv a^4\sum_x\bar\psi_L(x)\bar v_k^\dagger(x). 
\end{equation}
Basic requirements for the (ideal) basis vectors are 
\begin{equation}
   \hat P_-v_j=v_j,\qquad(v_j,v_k)=\delta_{jk}
\end{equation}
and 
\begin{equation}
   \bar v_kP_+=\bar v_k,\qquad(\bar v_j^\dagger,\bar v_k^\dagger)
   =\delta_{jk}
\end{equation}
so that
\begin{equation}
   \hat P_-\psi_L=\psi_L,\qquad\bar\psi_LP_+=\bar\psi_L.
\end{equation}

Let us consider how $\langle{\cal O}\rangle$ changes under the 
CP transformation of the gauge field, $U\to U^{\rm CP}$. The 
above framework gives 
\begin{equation}
   \left\langle{\cal O}\right\rangle(U^{\rm CP})
   =\int{\cal D}\psi_L^{\rm CP}{\cal D}\bar\psi_L^{\rm CP}\, 
{\cal O}^{\rm CP}\,
   \exp\left[\int\bar\psi_L^{\rm CP}D(U^{\rm CP})
\psi_L^{\rm CP}\right] 
\end{equation} 
and 
\begin{equation}
   {\cal D}\psi_L^{\rm CP}{\cal D}\bar\psi_L^{\rm CP}
   =\prod_j\rmd c_j^{\rm CP}\prod_k\rmd\bar c_k^{\rm CP}. 
\end{equation}
Here the expansion coefficients are defined by 
\begin{equation}
   \psi_L^{\rm CP}(x)=\sum_jv_j^{\rm CP}(x)c_j^{\rm CP},\qquad
   c_j^{\rm CP}=(v_j^{\rm CP},\psi_L^{\rm CP})
\end{equation}
and 
\begin{equation}
   \bar\psi_L^{\rm CP}(x)
   =\sum_k\bar c_k^{\rm CP}\bar v_k^{\rm CP}(x),\qquad
   \bar c_k^{\rm CP}=(\bar\psi_L^{{\rm CP}\dagger},
\bar v_k^{{\rm CP}\dagger}). 
\end{equation} 
The (ideal) basis vectors satisfy 
\begin{equation}
   \hat P_-(U^{\rm CP})v_j^{\rm CP}=v_j^{\rm CP},\qquad
   (v_j^{\rm CP},v_k^{\rm CP})=\delta_{jk}
\end{equation}
and
\begin{equation}
   \bar v_k^{\rm CP}P_+=\bar v_k^{\rm CP},\qquad
   (\bar v_j^{{\rm CP}\dagger},\bar v_k^{{\rm CP}\dagger}) 
=\delta_{jk}. 
\end{equation} 
In what follows, we take basis vectors as 
\begin{equation}
   \bar v_k^{\rm CP}=\bar v_k
\end{equation}
because both satisfy the {\it same\/} chirality constraint that 
is independent of gauge fields.

We thus examine the following modified substitution rule 
\begin{equation}
   \psi_L^{\rm CP}=-W^{-1}\left[\bar\psi_L{1\over\gamma_5
\Gamma_5(U)}\right]^T
   =-\sum_kW^{-1}\left[\bar v_k{1\over\gamma_5\Gamma_5(U)} 
\right]^T\bar c_k 
\end{equation} 
and 
\begin{equation}
   \bar\psi_L^{\rm CP}=[\gamma_5\Gamma_5(U)\psi_L]^TW
   =\sum_j[\gamma_5\Gamma_5(U)v_j]^TWc_j.
\end{equation}
This substitution is in fact consistent with the chirality 
constraint, $\hat P_-(U^{\rm CP})\psi_L^{\rm CP}
=\psi_L^{\rm CP}$ and~$\bar\psi_L^{\rm CP}P_+
=\bar\psi_L^{\rm CP}$. Moreover, the action takes the form 
identical to the original one under this substitution, as we 
already noted. The appearance of the singular factor 
$1/(\gamma_5\Gamma_5)$ is consistent with 
our ``no-go theorem''~\cite{fis1}. The question related to the 
existence of the inverse~$1/(\gamma_5\Gamma_5)$ is discussed 
later. These observations show that we should consider a change 
of integration variables from $(c_j^{\rm CP},\bar c_k^{\rm CP})$
  to~$(c_j,\bar c_k)$. These two sets are connected by 
\begin{equation}
   c_j^{\rm CP}=-\sum_ka^4\sum_x v_j^{{\rm CP}\dagger}(x)W^{-1}
   \left[\bar v_k(x){1\over\gamma_5\Gamma_5}\right]^T\bar c_k 
\end{equation} 
and 
\begin{equation}
\bar c_k^{\rm CP} =\sum_ja^4\sum_x[\gamma_5\Gamma_5v_j(x)]^TW
\bar v_k^\dagger(x) c_j. 
\end{equation} 
This transformation is however regular only if 
$\Tr\Gamma_5=n_+-n_-=0$, because 
\begin{equation}
   \hbox{\# of $c_j^{\rm CP}$}-\hbox{\# of $\bar c_k$}
   =\Tr\hat P_-(U^{\rm CP})-\Tr P_+
   =\Tr\hat P_+(U)-\Tr P_+
   =\Tr\Gamma_5
\end{equation}
and
\begin{equation}
   \hbox{\# of $\bar c_k^{\rm CP}$}-\hbox{\# of $c_j$}
   =\Tr\hat P_+-\Tr\hat P_-
   =\Tr\Gamma_5.
\end{equation}
So we assume $\Tr\Gamma_5=n_+-n_-=0$ in what follows; this is 
also necessary (though not sufficient) for the existence of the 
inverse of~$\gamma_5\Gamma_5$.

By defining
\begin{equation}
   \prod_j\rmd c_j^{\rm CP}\prod_k\rmd\bar c_k^{\rm CP}
   =J^{-1}\prod_j\rmd c_j\prod_k\rmd\bar c_k
\end{equation}
we have
\begin{eqnarray}
   J&=&\det\left\{-a^4\sum_x v_j^{{\rm CP}\dagger}(x)W^{-1}
   \left[\bar v_k(x){1\over\gamma_5\Gamma_5}\right]^T\right\}
   \det\left\{a^4\sum_x[\gamma_5\Gamma_5v_j(x)]^TW
   \bar v_k^\dagger(x)\right\}
\nonumber\\
   &=&\det\left\{a^4\sum_x[\gamma_5\Gamma_5v_j(x)]^TW
   \bar v_k^\dagger(x)\right\}
   \det\left[-a^4\sum_x
   \bar v_k(x){1\over\gamma_5\Gamma_5}(W^{-1})^Tv_j^{{\rm CP}*} 
(x)\right] \nonumber\\
   &=&\det\left\{-a^4\sum_x[\gamma_5\Gamma_5v_j(x)]^TW
   P_+{1\over\gamma_5\Gamma_5}(W^{-1})^Tv_k^{{\rm CP}*}(x) 
\right\} \nonumber\\
   &=&\det\left\{-a^4\sum_xv_j^{{\rm CP}\dagger}(x)W^{-1}
   \left[{1\over\gamma_5\Gamma_5(U)}\right]^TP_+^TW^T
   \gamma_5\Gamma_5(U)v_k(x)\right\}
\nonumber\\
   &=&\det\left[-a^4\sum_xv_j^{{\rm CP}\dagger}(x)
   {1\over\gamma_5\Gamma_5(U^{\rm CP})} \gamma_5\Gamma_5(U)
v_k(x)\right] \end{eqnarray} where we have used 
$\sum_k\bar v_k^\dagger(x)\bar v_k(y) =P_+\delta_{x,y}$ in 
deriving the third line, and~$W^T=W$ 
and~$P_-\gamma_5\Gamma_5=\gamma_5\Gamma_5\hat P_-$ in the forth 
line.

Clearly, whether the Jacobian~$J$ is unity or not depends on 
the  relation between $v_k$ and~$v_j^{\rm CP}$ which may be 
quite 
arbitrary (because these refer to different gauge fields, $U$ 
and~$U^{\rm CP}$, respectively). To investigate a minimal 
condition on $v_k$ and~$v_j^{\rm CP}$ such that $J=1$, we 
consider an infinitesimal variation of the gauge field specified
  by 
\begin{equation}
   \delta_\eta U(x,\mu)=a\eta_\mu(x)U(x,\mu).
\end{equation}
Under this variation, the Jacobian~$J=\det M$ changes 
as~$\delta_\eta\ln J=\tr\delta_\eta MM^{-1}$, where 
\begin{eqnarray}
   \delta_\eta M_{jk}
   &=&-a^4\sum_x\delta_\eta v_j^{{\rm CP}\dagger}(x)
   {1\over\gamma_5\Gamma_5(U^{\rm CP})}
\gamma_5\Gamma_5(U)v_k(x) \nonumber\\
   &&+a^4\sum_xv_j^{{\rm CP}\dagger}(x)
   {1\over\gamma_5\Gamma_5(U^{\rm CP})}\gamma_5
   \delta_\eta\Gamma_5(U^{\rm CP})
   {1\over\gamma_5\Gamma_5(U^{\rm CP})}
\gamma_5\Gamma_5(U)v_k(x) \nonumber\\
   &&-a^4\sum_xv_j^{{\rm CP}\dagger}(x)
   {1\over\gamma_5\Gamma_5(U^{\rm CP})}\gamma_5
\delta_\eta\Gamma_5(U)v_k(x)
\nonumber\\
   &&-a^4\sum_xv_j^{{\rm CP}\dagger}(x)
   {1\over\gamma_5\Gamma_5(U^{\rm CP})} \gamma_5\Gamma_5(U)
\delta_\eta v_k(x) 
\end{eqnarray} 
and 
\begin{equation}
   M_{kj}^{-1}
   =(v_k,{1\over\gamma_5\Gamma_5(U)}
   \gamma_5\Gamma_5(U^{\rm CP})v_j^{\rm CP}).
\end{equation}
Using $\sum_j v_j(x)v_j^\dagger(y)=\hat P_-(U)(x,y)$,
$\sum_j v_j^{\rm CP}(x)v_j^{{\rm CP}\dagger}(y)
=\hat P_-(U^{\rm CP})(x,y)$ 
and 
\begin{equation}
   {1\over\gamma_5\Gamma_5(U^{\rm CP})}
   \gamma_5\Gamma_5(U)\hat P_-(U)
   =\hat P_-(U^{\rm CP}){1\over\gamma_5\Gamma_5(U^{\rm CP})}
   \gamma_5\Gamma_5(U)
\end{equation}
together with  eq.~(2.45), we arrive at
\begin{equation}
   \delta_\eta\ln J=-i{\cal L}_\eta+i{\cal L}_\eta^{\rm CP}
   -\delta_\eta\Tr\Gamma_5(U)
\end{equation}
where ${\cal L}_\eta$ and~${\cal L}_\eta^{\rm CP}$ are 
so-called measure terms~\cite{luscher2}--\cite{luscher4}:
\begin{eqnarray}
   {\cal L}_\eta=i\sum_j(v_j,\delta_\eta v_j),\qquad
   {\cal L}_\eta^{\rm CP}=i\sum_j(v_j^{\rm CP},
\delta_\eta v_j^{\rm CP}) 
\end{eqnarray} 
which specify how the fermion path integral measure changes 
according to a change of gauge fields.

Recalling that $\Tr\Gamma_5$ is an integer which cannot change 
under an infinitesimal variation of the gauge field (or simply 
that we have set $\Tr\Gamma_5=0$), we see that the necessary 
condition for~$J=1$ is ${\cal L}_\eta^{\rm CP}={\cal L}_\eta$. 
Namely, for the CP invariance, the
(ideal) basis vectors have to be chosen such that
${\cal L}_\eta^{\rm CP}={\cal L}_\eta$. Conversely, if
${\cal L}_\eta^{\rm CP}={\cal L}_\eta$, we see that the 
Jacobian~$J$ is a constant which can depend only on the 
topological properties of each sector. In the vacuum sector, in 
which the vacuum $U_0=1$ is contained, we can determine this 
constant and obtain $J=1$ because $U_0^{\rm CP}=1=U_0$. So, for 
the vacuum sector, ${\cal L}_\eta^{\rm CP}={\cal L}_\eta$ 
implies $J=1$.

In our previous work, we have shown that the conditions on the 
ideal measure factor (which appear in the reconstruction 
theorem of chiral gauge
theory~\cite{luscher2,luscher3}) are consistent with the choice 
 ${\cal L}_\eta^{\rm CP}={\cal L}_\eta$~\cite{fis2}. The unit 
Jacobian condition (in the vacuum sector) is thus equivalent to 
the existence of the ideal measure factor in this sense.

In fact, the CP invariance in the sense that we can ignore the 
Jacobian associated with the above modified CP transformation 
is shown more generally, when there is no modes such that 
$\gamma_5\Gamma_5\Psi(x)=0$, namely, $N_+=N_-=0$. In this case, 
one can show that the Jacobian is a pure-phase, $J=e^{i\theta}$.
  With the CP invariant choice of the fermion measure terms, 
${\cal L}_\eta^{\rm CP}={\cal L}_\eta$, the phase~$\theta$ is a 
constant depending only on the topological sector, as we have 
shown above. Such a constant breaking of CP, however, may be 
re-absorbed into the basis vectors $v_j$ and~$v_j^{\rm CP}$ 
(this operation does not change the measure terms), or 
equivalently may be absorbed into the phase 
factor~$\vartheta_M$ for each topological sector. 
This apparent CP breaking is thus harmless. This is completely 
consistent with our result in the previous work~\cite{fis2} 
where the CP invariance of path integral (in the topologically 
trivial 
sector) except for propagators is shown.
We present the proof of the above statement below.

\noindent
{\it Proof of $|J|^2=1$}:

\noindent
Our Jacobian factor $J$ is expressed as 
\begin{equation}
J=\det\left[-a^4\sum_xv_j^{{\rm CP}\dagger}(x) {1\over\gamma_5
\Gamma_5(U^{\rm CP})}t_k(x)\right] 
\det\left[a^4\sum_xt_j^\dagger(x)\gamma_5\Gamma_5(U)v_k(x)\right]
\end{equation}
where $\{t_j(x)\}$ is any orthonormal complete set of vectors 
such that $P_-t_j=t_j$. First, one can easily see that $|J|^2$ 
is invariant under a unitary transformation of bases, 
$v_j^{\rm CP}$ and~$v_j$. We may therefore choose any bases (as 
long as they are consistent with the chirality
constraints) in evaluating~$|J|^2$. A convenient choice is 
the ``auxiliary basis'' defined by $H$: 
\begin{equation}
   v_j(x)=w_j(x)=\hat P_-u_j(x),\qquad(w_j,w_k)=\delta_{jk}
\end{equation}
or more explicitly, 
\begin{equation}
   w_j=\varphi_0^-
\end{equation}
which satisfies $H^2w_j=0$, and
\begin{equation}
   w_j={1\over\sqrt{2[1-\lambda_jf(\lambda_j^2)]}}
   \left\{\sqrt{1-\lambda_j^2f^2(\lambda_j^2)}\varphi_j
   -[1-\lambda_jf(\lambda_j^2)]\widetilde\varphi_j\right\}
\end{equation}
which satisfies $H^2w_j=\lambda_j^2w_j$. (We use basically half 
of the eigenstates of $H$.) See~Appendix for notational 
conventions. Similarly we set \begin{equation}
   v_j^{\rm CP}(x)=w_j^{\rm CP}(x)
   =\hat P_-(U^{\rm CP})u_j^{\rm CP}(x),\qquad
   (w_j^{\rm CP},w_k^{\rm CP})=\delta_{jk}
\end{equation}
where $u_j^{\rm CP}(x)$ is the eigenfunction 
of~$H^2(U^{\rm CP})$, $H^2(U^{\rm CP})u_j^{\rm CP}(x) 
={\lambda_j^{\rm CP}}^2u_j^{\rm CP}(x)$. We also use 
\begin{equation}
   t_j(x)=P_-u_j(x),\qquad (t_j,t_k)=\delta_{jk}
\end{equation}
namely,
\begin{equation}
   t_j=\varphi_0^-
\end{equation}
and
\begin{equation}
   t_j={1\over\sqrt{2[1-\lambda_jf(\lambda_j^2)]}}
   \left\{\sqrt{1-\lambda_j^2f^2(\lambda_j^2)}\varphi_j
   -[1+\lambda_jf(\lambda_j^2)]\widetilde\varphi_j\right\}.
\end{equation}

When there is no modes such that $\gamma_5\Gamma_5\Psi(x)=0$, 
the above vectors span complete sets in the space restricted by 
the chirality constraints. Then, using properties of~$\Gamma_5$,
  it is straightforward to see that 
\begin{equation}
   a^4\sum_xt_j^\dagger(x)\gamma_5\Gamma_5(U)v_k(x)
   =\sqrt{1-\lambda_j^2f^2(\lambda_j^2)}\delta_{jk}
\end{equation}
and
\begin{equation}
   a^4\sum_xv_j^{{\rm CP}\dagger}(x)
   {1\over\gamma_5\Gamma_5(U^{\rm CP})}t_k(x)
   ={1\over\sqrt{1-{\lambda_j^{\rm CP}}^2f^2
   ({\lambda_j^{\rm CP}}^2)}}\delta_{jk}.
\end{equation}
Therefore, we have
\begin{equation}
 |J|^2=
{\prod_j[1-\lambda_j^2f^2(\lambda_j^2)]\over
\prod_k[1-{\lambda_k^{\rm CP}}^2f^2({\lambda_k^{\rm CP}}^2)]}. 
\end{equation} 
This combination is, however, unity because for 
$\lambda_j\neq0$ (and for~$\lambda_j\neq\Lambda$, which is our 
assumption), the eigenvalues are degenerate as 
$\lambda_j^2={\lambda_j^{\rm CP}}^2$, as one can confirm by 
using the relations in Appendix and~eq.~(2.45).

\section{CP (or C) transformation and Yukawa couplings}

The CP symmetry is of course broken in the presence of the Higgs
  coupling in chiral gauge theory. For example,\footnote{We 
assume that the left-handed fermion~$\psi_L(x)$ belongs to the 
representation~$R_L$ of the gauge group and the right-handed 
fermion~$\psi_R(x)$ belongs to~$R_R$ (the Higgs field~$\phi(x)$ 
transforms as~$R_L\otimes(R_R)^*$). The gauge couplings in the 
Dirac operators~$D(U_1)$ and $D(U_2)$, and correspondingly in 
$\hat P_-(U_1)$ and~$\hat P_+(U_2)$, are thus defined with 
respect to the representations~$R_L$ and $R_R$, respectively.} 
\begin{eqnarray}
   {\cal L}&=&\bar\psi_LD(U_1)\psi_L+\bar\psi_RD(U_2)\psi_R
   +2g(\bar\psi_L\phi\psi_R+\bar\psi_R\phi^\dagger\psi_L)
\nonumber\\
&=&\bar\psi P_+D(U_1)\hat P_-(U_1)\psi+\bar\psi P_-D(U_2)
\hat P_+(U_2)\psi \nonumber\\
   &&+2g\left[\bar\psi P_+\phi\hat P_+(U_2)\psi
   +\bar\psi P_-\phi^\dagger\hat P_-(U_1)\psi\right] 
\end{eqnarray} 
where CP is broken not only in the kinetic term but also in the 
Higgs couplings. Under the CP transformation, 
\begin{eqnarray}
&&U_1\to U_1^{\rm CP},\qquad U_2\to U_2^{\rm CP}
\nonumber\\ &&\bar\psi\to\psi^TW,\qquad\psi\to-W^{-1}\bar\psi^T,
\nonumber\\
&&\phi\to\phi^*
\end{eqnarray}
this Lagrangian is transformed to
\begin{eqnarray}
{\cal L}^{\rm CP}&=&
\bar\psi\gamma_5\hat P_+(U_1)\gamma_5D(U_1)P_-\psi
+\bar\psi\gamma_5\hat P_-(U_2)\gamma_5D(U_2)P_+\psi \nonumber\\
&&+2g\left[\bar\psi\gamma_5\hat P_+(U_1)\gamma_5\phi P_+\psi
+\bar\psi\gamma_5\hat P_-(U_2)\gamma_5
\phi^\dagger P_-\psi\right]. 
\end{eqnarray} 
This is again interpreted as a change of representation of 
chiral projection operators, from $P_\pm$ and~$\hat P_\pm$ 
to $\gamma_5\hat P_\pm\gamma_5$ and~$P_\pm$, constructed from 
a vector-like Ginsparg-Wilson theory if one introduces two sets 
of fermion fields $\psi^{(1)}$ and~$\psi^{(2)}$ 
in~eq.~(4.1)
\begin{eqnarray}
&&\psi_L=\hat P_-(U_1)\psi^{(1)}, \qquad\bar\psi_L
=\bar\psi^{(1)}P_+,
\nonumber\\
&&\psi_R=\hat P_+(U_2)\psi^{(2)}, \qquad\bar\psi_R
=\bar\psi^{(2)}P_-.
\end{eqnarray}
In a perturbative treatment of the Higgs coupling, the analysis 
of CP symmetry becomes identical to that of the pure chiral 
gauge theory, as was shown elsewhere~\cite{fis2}. For a 
non-perturbative treatment of the Higgs coupling but in the 
topologically trivial sector, one can use the modified CP 
transformation motivated by the domain wall fermion, 
\begin{eqnarray}
&&\psi_L\to\psi_L^{\rm CP}=-W^{-1} \left[\bar\psi_L
\frac{1}{\gamma_5\Gamma_5(U_1)}\right]^T,
\nonumber\\
&&\bar\psi_L\to\bar\psi_L^{\rm CP}
= [\gamma_5\Gamma_5(U_1)\psi_L]^TW, \nonumber\\ 
&&\psi_R\to\psi_R^{\rm CP} =-W^{-1}\left[\bar\psi_R
\frac{1}{\gamma_5\Gamma_5(U_2)}\right]^T,
\nonumber\\
&&\bar\psi_R\to\bar\psi_R^{\rm CP} 
=[\gamma_5\Gamma_5(U_2)\psi_R]^TW 
\end{eqnarray} 
which keeps the action~(4.1) invariant. The invariance of the 
Higgs coupling is confirmed by noting, for example, 
\begin{eqnarray}
\bar\psi_L^{\rm CP}P_+\phi^{CP}\hat P_+(U_2^{\rm CP}) 
\psi_R^{\rm CP}
&=&-\left[\gamma_5\Gamma_5(U_1)\psi_L\right]^TWP_+\phi^*
P_+\gamma_5\Gamma_5(U_2^{\rm CP})W^{-1} 
\left[\bar\psi_R\frac{1}{\gamma_5\Gamma_5(U_2)}\right]^T
\nonumber\\ &=&-[\gamma_5\Gamma_5(U_1)\psi_L]^TWP_+
\phi^*P_+W^{-1}
\bar\psi_R^T
\nonumber\\
&=&\bar\psi_RP_-\phi^\dagger P_-\gamma_5\Gamma_5(U_1)\psi_L 
\nonumber\\
&=&\bar\psi_RP_-\phi^\dagger\hat P_-(U_1)\psi_L 
\end{eqnarray} 
where we used $P_+\hat P_+(U_2^{\rm CP}) 
=P_+P_+\gamma_5\Gamma_5(U_2^{\rm CP})$ and 
$P_-P_-\gamma_5\Gamma_5(U_1)=P_-\hat P_-(U_1)$. We can thus 
repeat the analysis of the previous section and confirm that 
the path integral is invariant under the modified CP 
transformation except for the propagators which are determined 
by the source terms. The essence of CP analysis in the domain 
wall representation is included in this analysis.

It would be interesting if one can generally establish the CP 
invariance except for the propagators 
\begin{eqnarray} 
&&\left\langle\psi_L(x)\bar\psi_L(y)\right\rangle
=\hat P_-(U_1)\frac{1}{D(U_1) -2g\phi\frac{1}{D(U_2)}
2g\phi^\dagger}P_+,
\nonumber\\
&&\left\langle\psi_L(x)\bar\psi_R(y)\right\rangle
=-\hat P_-(U_1)\frac{1}{D(U_1) -2g\phi\frac{1}{D(U_2)}
2g\phi^\dagger}
2g\phi\frac{1}{D(U_2)}P_-,
\nonumber\\
&&\left\langle\psi_R(x)\bar\psi_R(y)\right\rangle
=\hat P_+(U_2)\frac{1}{D(U_2) -2g\phi^\dagger\frac{1}{D(U_1)}
2g\phi}P_-,
\nonumber\\
&&\left\langle\psi_R(x)\bar\psi_L(y)\right\rangle
=-\hat P_+(U_2)\frac{1}{D(U_2) -2g\phi^\dagger\frac{1}{D(U_1)}
2g\phi}
2g\phi^\dagger\frac{1}{D(U_1)}P_+,
\end{eqnarray}
which depend on the specific choice of chiral projection 
operators $P_{\pm}$ and $\hat{P}_{\pm}$ as in (4.7) (or 
$\gamma_{5}\hat{P}_{\pm}\gamma_{5}$ and $P_{\pm}$ after CP 
transformation), after 
summing over the topological sectors but 
without using the explicit diagonal representation of the 
action (which was used in our previous paper~\cite{fis2}).
  
It is shown that CP is broken even in the vector-like theory in 
the presence of chiral symmetric Yukawa couplings. For example, 
one may consider a theory with Abelian flavor symmetry (by using
  $P_\pm\hat P_\pm=P_\pm\gamma_5\Gamma_5$) 
\begin{eqnarray}
{\cal L}&=&\bar\psi_RD\psi_R+\bar\psi_LD\psi_L 
-m\left(\bar\psi_R\psi_L+\bar\psi_L\psi_R\right)
+2g\left(\bar\psi_L\phi\psi_R+\bar\psi_R\phi^\dagger\psi_L\right)
\nonumber\\
&=&\bar\psi D\psi-m\bar\psi\gamma_5\Gamma_5\psi
+2g\bar\psi\left(P_+\phi\hat P_++P_-\phi^\dagger\hat P_-\right)
\psi \nonumber\\
&=&\bar\psi D\psi-m\bar\psi\gamma_5\Gamma_5\psi
+2g\bar\psi\left(P_+\phi P_++P_-\phi^\dagger P_-\right)
\gamma_5\Gamma_5\psi. 
\end{eqnarray} 
The Yukawa coupling in this Lagrangian is not invariant under 
CP transformation\footnote{Under parity we have $\phi\to\phi^*$,
  and thus under CP we have $\phi\to\phi^*$.} 
\begin{eqnarray}
   &&\bar\psi\to\psi^TW,\qquad\psi\to-W^{-1}\bar\psi^T,
\nonumber\\ 
   &&WD(U^{\rm CP})W^{-1}=D(U)^T,\qquad
   W\gamma_5\Gamma_5(U^{\rm CP})W^{-1}=[\gamma_5\Gamma_5(U)]^T,
\nonumber\\
   &&W\phi W^{-1}=\phi^*.
\end{eqnarray}
This non-invariance arises from 
\begin{equation}
   [\gamma_5,\gamma_5\Gamma_5]\neq0,\qquad
   [\phi(x),\gamma_5\Gamma_5]\neq0.
\end{equation}
For a real constant $\phi$, these conditions are cleared, 
and the Yukawa coupling is reduced to the mass term. The above 
CP non-invariance is of course interpreted as a change from one 
representation of lattice chiral projectors to another, just as 
we discussed in the case of pure chiral gauge theory.

One can re-write the above Lagrangian in terms of the domain 
wall fermion as 
\begin{eqnarray}
   {\cal L}&=&\bar qD\frac{1}{\gamma_5\Gamma_5}q-m\bar qq
   +2g\bar q\left(P_+\phi P_++P_-\phi^\dagger P_-\right)q
   +\bar Q\frac{1}{\gamma_5\Gamma_5}Q
\end{eqnarray}
which is {\it invariant\/} under CP transformation 
\begin{eqnarray}
   &&\bar q\to q^TW,\qquad q\to-W^{-1}\bar q^T,
\nonumber\\ 
   &&\bar Q\to Q^TW,\qquad Q\to-W^{-1}\bar Q^T,
\nonumber\\
   &&WD(U^{\rm CP})W^{-1}=D(U)^T,\qquad
   W\gamma_5\Gamma_5(U^{\rm CP})W^{-1}=[\gamma_5\Gamma_5(U)]^T,
\nonumber\\
   &&W\phi W^{-1}=\phi^*
\end{eqnarray}
if one notes that the overall signature of the last term 
in~eq.~(4.11) is immaterial.

To see the breaking of CP symmetry in this context of the 
domain wall representation, we introduce the source terms for 
the fermion fields which specifies general correlation 
functions. For the local Ginsparg-Wilson variables, we have 
\begin{equation}
\int{\cal L}_{\rm source}=\int\left(\bar\psi\eta
+\bar\eta\psi\right)
\end{equation}
which is invariant under CP transformation
\begin{eqnarray}
   &&\bar\psi\to\psi^TW,\qquad\psi\to-W^{-1}\bar\psi^T,
\nonumber\\
   &&\bar\eta\to\eta^T_wW,\qquad\eta\to-W^{-1}\bar\eta_w^T.
\end{eqnarray}
The source terms are translated in the language of the domain 
wall fermion as 
\begin{equation}
\int{\cal L}_{\rm source}
=\int\left(\bar q\eta+\bar\eta\frac{1}{\gamma_5\Gamma_5}q\right)
\end{equation}
which is transformed under CP symmetry
\begin{equation}
   \bar q\to q^TW,\qquad q\to-W^{-1}\bar q^T
\end{equation}
to
\begin{equation}
\int\left(\bar q\frac{1}{\gamma_5\Gamma_5}\eta_w
+\bar\eta_wq\right).
\end{equation}
To recover the original source terms,\footnote{This 
complication does not appear to be resolved by an argument of 
the use of equations of motion for external field lines in the 
non-perturbative treatment of the Yukawa coupling. 
See~ref.~\cite{fis2}.} we need to perform the re-definition of 
field variables 
\begin{equation}
q\to\frac{1}{\gamma_5\Gamma_5}q,\qquad\bar q\to\bar q 
\gamma_5\Gamma_5 
\end{equation} 
but the Yukawa coupling is not invariant under this 
re-definition because of 
$[\gamma_5,\gamma_5\Gamma_5]\neq0$ and 
$[\phi(x),\gamma_5\Gamma_5]\neq0$. The propagator is thus 
modified under CP as 
\begin{eqnarray}
   &&\frac{1}{D/(\gamma_5\Gamma_5)-m+2g(P_+\phi P_+
+P_-\phi^\dagger P_-)}
   \times\frac{1}{\gamma_5\Gamma_5}
\nonumber\\
&&\neq\frac{1}{\gamma_5\Gamma_5}\times
\frac{1}{D/(\gamma_5\Gamma_5)-m+2g(P_+\phi P_+
+P_-\phi^\dagger P_-)}.
\end{eqnarray}

We arrive at the same conclusion by using the fermionic 
representation of the domain wall fermion with a chiral 
symmetric Yukawa coupling 
\begin{eqnarray}
   {\cal L}&=&\bar qD\frac{1}{\gamma_5\Gamma_5}q-m\bar qq
   +2g\bar q(P_+\phi P_++P_-\phi^\dagger P_-)q
   +\bar S\gamma_5\Gamma_5S
\end{eqnarray}
if one  assigns CP transformation
\begin{eqnarray}
   \bar S\to S^TW,\qquad S\to-W^{-1}\bar S^T.
\end{eqnarray}
The action is invariant under CP transformation, but to keep 
the source terms invariant one needs to perform a field 
re-definition which is not compatible with the Yukawa coupling.

\section{Majorana fermion}

The above complication of CP symmetry (or equivalently charge 
conjugation symmetry since the parity is normal in  the above 
model) for the vector-like theory with the chiral invariant 
Yukawa coupling gives rise to a difficulty in defining Majorana 
fermions in a Euclidean sense~\cite{fi1,fis1}. 
Following the standard procedure, we replace the field 
variables~\cite{nicolai, van nieuwenhuizen,suzuki3} 
\begin{eqnarray}
&&\psi=(\chi+i\eta)/\sqrt2,
\nonumber\\
&&\bar\psi=(\chi^TC-i\eta^TC)/\sqrt2
\end{eqnarray}
in the Lagrangian written in the Ginsparg-Wilson fermions. We 
naively expect\footnote{If $(CO)^T=-CO$ or equivalently 
$COC^{-1}=O^T$ for a general operator~$O$, the cross term 
vanishes $\eta^TCO\chi-\chi^TCO\eta=0$ by using the 
anti-commuting property of $\chi$ and~$\eta$. In the presence 
of background gauge field, we assume that the representation of 
gauge symmetry is real.} 
\begin{eqnarray}
   {\cal L}&=&\frac{1}{2}\chi^TCD\chi-\frac{1}{2}m\chi^TC
\gamma_5\Gamma_5\chi
+g\chi^TC\left(P_+\phi\hat P_++P_-\phi^\dagger\hat P_-\right)
\chi \nonumber\\
&&+\frac{1}{2}\eta^TCD\eta-\frac{1}{2}m\eta^TC
\gamma_5\Gamma_5\eta
+g\eta^TC\left(P_+\phi\hat P_++P_-\phi^\dagger\hat P_-\right)
\eta. 
\end{eqnarray} 
One would then define the Majorana fermion $\chi$ (or $\eta$) 
and the resulting Pfaffian. But this actually fails since the 
cross terms between $\chi$ and~$\eta$ do not quite vanish due 
to the complications in the charge conjugation.

If one uses the domain wall fermion with ``fermionic'' 
variables, one may make the 
replacement\footnote{The Majorana reduction of the bosonic 
fermion~$Q$ in the conventional domain wall fermion is 
non-trivial, since $\int\lambda^TC\frac{1}{\gamma_5\Gamma_5}
\lambda=0$ for a bosonic spinor.} 
\begin{eqnarray}
&&q=(\chi+i\eta)/\sqrt2,\nonumber\\
&&\bar q=\left(\chi^TC-i\eta^TC\right)/\sqrt2,
\nonumber\\
&&S=(\lambda+i\rho)/\sqrt2,
\nonumber\\
&&\bar S=\left(\lambda^TC-i\rho^TC\right)/\sqrt2.
\end{eqnarray}
One can then define the Majorana fermions $\chi$ or~$\eta$ (and 
$\lambda$ or~$\rho$) by
\begin{eqnarray}
   {\cal L}&=&\frac{1}{2}\chi^TCD\frac{1}{\gamma_5\Gamma_5}\chi 
   -\frac{1}{2}m\chi^TC\chi
   +g\chi^TC\left(P_+\phi P_++P_-\phi^\dagger P_-\right)
\chi \nonumber\\
   &&+\frac{1}{2}\eta^TCD\frac{1}{\gamma_5\Gamma_5}\eta
   -\frac{1}{2}m\eta^TC\eta
   +g\eta^TC\left(P_+\phi P_++P_-\phi^\dagger P_-\right)
\eta \nonumber\\
   &&+\frac{1}{2}\lambda^TC\gamma_5\Gamma_5\lambda
   +\frac{1}{2}\rho^TC\gamma_5\Gamma_5\rho,
\end{eqnarray}
namely, one may define a Majorana fermion by
\begin{eqnarray}
{\cal L}_M&=&\frac{1}{2}\chi^TCD\frac{1}{\gamma_5\Gamma_5}\chi 
-\frac{1}{2}m\chi^TC\chi
+g\chi^TC\left(P_+\phi P_++P_-\phi^\dagger P_-\right)\chi 
\nonumber\\
&&+\frac{1}{2}\lambda^TC\gamma_5\Gamma_5\lambda.
\end{eqnarray}
This theory is however non-local due to the singularities 
in~$1/(\gamma_5\Gamma_5)$.

In the level of path integral, one may modify the above 
Lagrangian by writing 
\begin{eqnarray}
&&\int{\cal D}\chi{\cal D}\lambda\,\exp\left(\int{\cal L}_M
\right) \nonumber\\
&&=\int{\cal D}\chi\,
\exp\biggl\{\int\biggl[\frac{1}{2}\chi^TCD\chi
-\frac{1}{2}m\chi^TC\gamma_5\Gamma_5\chi
\nonumber\\
   &&\qquad\qquad\qquad\qquad
   +g\chi^TC\sqrt{\gamma_5\Gamma_5}
\left(P_+\phi P_++P_-\phi^\dagger P_-\right)
\sqrt{\gamma_5\Gamma_5}\chi
   \biggr]\biggr\}
\end{eqnarray}
where we made a formal rescaling 
\begin{equation}
   \chi\to\sqrt{\gamma_5\Gamma_5}\chi,\qquad
   \sqrt{\gamma_5\Gamma_5}\lambda\to\lambda.
\end{equation}
This rescaling formally removes the singular factor 
$1/(\gamma_5\Gamma_5)$ and makes the auxiliary 
fermion~$\lambda$ decouple. This final path integral is however 
not what we expect for the Ginsparg-Wilson fermion because of 
$[\gamma_5,\gamma_5\Gamma_5]\neq0$ 
and~$[\phi(x), \gamma_5\Gamma_5]\neq0$, which caused the 
failure of the charge conjugation symmetry. As for the Pfaffian 
and the determinant factor without the external fermion lines, 
one may adopt the above definition of the Majorana fermion, 
which is consistent up to a possible non-locality arising 
from~$\sqrt{\gamma_5\Gamma_5}$.

A difficulty in defining the Majorana fermion is clearly seen 
when one considers the source terms for the Ginsparg-Wilson 
fermion, as we did in the analysis of CP symmetry, 
\begin{eqnarray}
   \int\left(\bar J\psi+\bar\psi J\right)
   =\int\left(\chi^TCJ_1+\eta^TCJ_2\right)
\end{eqnarray}
where the Majorana sources are defined by
\begin{eqnarray}
J=(J_1+iJ_2)/\sqrt2,
\qquad\bar J=\left(J_1^TC-iJ_2^TC\right)/\sqrt2.
\end{eqnarray}
The derivatives with respect to the source $J_1$ give rise to 
correlation functions of the would-be Majorana fermion~$\chi$, 
which we failed to define for the Ginsparg-Wilson fermion.

The corresponding source terms for the domain wall fermion are 
given by 
\begin{eqnarray}
 &&\int\left(\bar J\frac{1}{\gamma_5\Gamma_5}q+\bar qJ\right) 
\nonumber\\
   &&=\int\frac{1}{2}
   \left\{\left[\left(1+\frac{1}{\gamma_5\Gamma_5}\right)
\chi\right]^TC
-i\left[\left(1-\frac{1}{\gamma_5\Gamma_5}\right)\eta\right]^TC
   \right\}J_1
\nonumber\\
   &&+\int\frac{1}{2}
 \left\{\left[\left(1+\frac{1}{\gamma_5\Gamma_5}\right)
\eta\right]^TC
 +i\left[\left(1-\frac{1}{\gamma_5\Gamma_5}\right)\chi\right]^TC
   \right\}J_2
\end{eqnarray}
where we used the variables supposed to describe Majorana 
fermions in the domain wall representation (5.4). This 
expression of source terms shows that neither of the Majorana 
fermions, $\chi$ and~$\eta$, defined by the domain wall fermion 
correspond to the Majorana fermion generated by the 
source~$J_1$, for example. Besides, the correlation functions 
generated by differentiating with respect to~$J_1$ contain the 
species-doubler poles in~$1/(\gamma_5\Gamma_5)$. This shows 
that we cannot define the Majorana fermion consistently for 
physical processes in the presence of the chiral symmetric 
Yukawa coupling. The conflict among chiral symmetry, strict 
locality and Majorana condition persists. The condition for the 
presence of Majorana fermions is in a sense more demanding than 
the CP invariance. The Majorana fermion requires a Lagrangian 
self-symmetric under charge conjugation, while CP symmetry 
requires the invariance of the path integral after summing over 
all topological sectors.

In a supersymmetric Wess-Zumino model on the lattice, one needs 
to define the constraint-free Majorana fermion.\footnote{If one
uses the Weyl fermion defined by the Ginsparg-Wilson operator,
the constant spinor parameter appearing in supersymmetry
transformation is constrained by projection operators. This
leads to complications, in particular, in the presence of the
background gauge field.}
A past attempt to define the Wess-Zumino model is given 
by~\cite{fi1,fujikawa2} 
\begin{eqnarray}
{\cal L}_{\rm WZ}&=&\frac{1}{2}\chi^TC
\frac{1}{\gamma_5\Gamma_5}D\chi
   -\phi^\dagger D^\dagger D\phi
   +F^\dagger\frac{1}{\Gamma_5^2}F
   +\frac{1}{2}m\chi^TC\chi+m\left[F\phi+(F\phi)^\dagger\right]
\nonumber\\
   &&+g\chi^TC\left(P_+\phi P_++P_-\phi^\dagger P_-\right)\chi
   +g\left[F\phi^2+(F\phi^2)^\dagger\right]
\end{eqnarray}
where $\phi$ stands for the complex scalar field and $F$ for the
 auxiliary field. The operator~$D$ is the (free) Ginsparg-Wilson
 operator, and when $D^\dagger D$ appears in the bosonic sector 
we adopt the convention to discard the (unit) Dirac matrix. The 
Majorana fermion~$\chi$ and its Yukawa couplings are the same as
 those we find for the above domain wall representation. 
However, a crucial difference is that the Pauli-Villars 
field~$S$ is now replaced by the ``physical'' field~$F$. 
For this reason, we regard the field~$\chi$ 
in~${\cal L}_{\rm WZ}$ as a primary definition of the Majorana 
fermion, though it is defined by a non-local Lagrangian. 
The singular factor~$1/(\gamma_5\Gamma_5)$ for the Majorana 
fermion is canceled by the same factor coming from the 
auxiliary field~$F^\dagger[1/(\Gamma^2_5)]F$. In fact one can 
confirm that the free part of~${\cal L}_{\rm WZ}$ is invariant 
under a lattice version of supersymmetry~\cite{fujikawa2} 
\begin{eqnarray}
   &&\delta\chi=-\Gamma_5\frac{1}{a}H(A-i\gamma_5B)\epsilon
   -(F-i\gamma_5G)\epsilon,
\nonumber\\
   &&\delta A=\epsilon^TC\chi=\chi^TC\epsilon,
\nonumber\\
 &&\delta B=-i\epsilon^TC\gamma_5\chi=-i\chi^TC\gamma_5\epsilon,
\nonumber\\
   &&\delta F=\epsilon^TC\Gamma_5\frac{1}{a}H\chi,
\nonumber\\
   &&\delta G=i\epsilon^TC\Gamma_5\frac{1}{a}H\gamma_5\chi
\end{eqnarray}
with a constant Majorana-type Grassmann parameter~$\epsilon$. 
Here we defined 
\begin{equation}
\phi\to\frac{1}{\sqrt2}(A+iB),
\qquad F\to\frac{1}{\sqrt2}(F-iG) 
\end{equation} 
and $H=a\gamma_5D$. This construction of~${\cal L}_{\rm WZ}$ is 
not completely satisfactory, but it may be amusing to see that a
 certain aspect of the domain wall fermion may play an essential
 role in the construction of Majorana fermions.

\section{Discussion}

We have examined the CP properties of a domain wall fermion 
where light field variables $q$ and~$\bar q$ and the 
Pauli-Villars fields $Q$ and~$\bar Q$ are used. It was first 
shown that the variables $q_{L}$ and~$\bar q_{L}$ cannot 
describe the topological properties, and the full physical 
contents are only described by the local Ginsparg-Wilson 
variables $\psi_{L}$ and~$\bar\psi_{L}$. The domain wall 
variables $q$ and~$\bar q$ in the infinite flavor limit, which 
themselves exhibit nice CP and 
charge conjugation properties, cannot help to resolve the 
difficulty associated with  CP symmetry in chiral gauge 
theory~\cite{hasenfratz} and the failure of the Majorana 
condition in the presence of chiral symmetric Yukawa 
couplings~\cite{fi1}.
 
The conflict among the good chiral property, strict locality and
 CP (or charge conjugation) symmetry thus persists. The CP 
transformation sends one representation of lattice chiral gauge 
theory into another representation of lattice chiral gauge 
theory,
 which are constructed from the same vector-like theory defined 
by the Ginsparg-Wilson operator $D$. The violation of CP 
symmetry in the Lagrangian level is partly resolved by summing 
over various topological sectors~\cite{fis2}, and the CP 
non-invariance is manifested by the change of propagators. In 
the presence of Higgs couplings, the complications with CP 
symmetry become more involved since the chiral projection 
operators are determined by the Ginsparg-Wilson operator which 
depends only on the gauge field whereas the non-perturbative 
fermion propagator contains Higgs couplings as well. As for a 
definition of Majorana fermions in the presence of chiral 
symmetric Yukawa couplings, an action which is symmetric under 
the charge conjugation is required. The Ginsparg-Wilson fermions
 cannot be used in this context. As a tentative (and not 
complete) resolution of this conflict, we mentioned a use of the
 domain wall-like representation for the supersymmetric 
Wess-Zumino model where the auxiliary field~$F$ plays a role of 
the Pauli-Villars fields.

We have analyzed only the infinite flavor limit $N\to\infty$ in 
the domain wall fermion, where chiral symmetry is well-defined. 
It will be interesting to examine if the above conflict is 
already seen for the finite~$N$ domain wall fermion where 
the operator~$D_N/(1-aD_N)$ is local (see Ref.~\cite{kikukawa4} 
for the locality of $D_{N}$), though 
precise chiral symmetry is not defined.\footnote{One may, for 
example, argue that the domain wall variables $q_{L}$ 
and~$\bar q_{L}$, which become non-local and cannot 
describe topological properties in the limit~$N=\infty$, are not
 the suitable variables to describe physical correlation 
functions even for finite $N$, to the extent that the finite 
$N$ theory is intended to be an approximation to the theory with
 $N=\infty$.}

Our analysis of various complications is based on the singular 
behavior of 
\begin{equation}
   \frac{1}{\gamma_5\Gamma_5}=\frac{1}{1-aD(\gamma_5aD)^{2k}}
\end{equation}
in the context of general Ginsparg-Wilson operators. This 
factor contains poles at the positions of the would-be species 
doublers which have a mass~$1/a$ in the case of free fermions, 
and topological poles in the presence of instantons. This mass 
value approaches~$\infty$ in the limit~$a\to0$, and those 
particles are naively expected to decouple from the Hilbert 
space in the same limit. The singularity at~$1/a$ causes 
non-locality in a strict sense and thus cannot be consistent in 
all respects~\cite{chiu2}, but one might hope that the 
singularity may not be so serious in a suitable limit~$a\to0$ 
in some practical applications. This issue may deserve further 
analyses and, in any case, would lead to a better understanding 
of the domain wall fermion.

\noindent
{\bf Acknowledgments}

One of us (KF) thanks H.~Neuberger for a useful conversation on 
the domain wall fermion. He also thanks S.~Aoki, Y.~Kikukawa and
 J.~Kubo for discussions at Summer Institute 2002 in 
Fuji-Yoshida, and all the members of C.N.~Yang Institute for 
Theoretical Physics, Stony Brook, for their hospitality. The 
other of us (HS) thanks Y. Kikukawa and Y. Taniguchi for 
discussions.

\appendix

\section{Representation of the Ginsparg-Wilson algebra}

We here summarize the representation of the general 
Ginsparg-Wilson relation~\cite{fis1,fis2} 
$H\gamma_5+\gamma_5H=2H^2f(H^2)$. Let us consider the 
eigenvalue problem 
\begin{equation}
   H\varphi_n(x)=\lambda_n\varphi_n(x),\qquad
   (\varphi_n,\varphi_m)=\delta_{nm}.
\end{equation}
We first note $H\Gamma_5\varphi_n(x)=-\Gamma_5H\varphi_n(x)
=-\lambda_n\Gamma_5\varphi_n(x)$,
and
\begin{equation}
   (\Gamma_5\varphi_n,\Gamma_5\varphi_m)
   =[1-\lambda_n^2f^2(\lambda_n^2)]\delta_{nm}.
\end{equation}
These relations show that eigenfunctions with 
$\lambda_n\neq0$ and~$\lambda_nf(\lambda_n^2)\neq\pm1$ come in 
pairs as $\lambda_n$ and~$-\lambda_n$ (when $\lambda_n=0$, 
$\varphi_0(x)$ and~$\Gamma_5\varphi_0(x)$ are not necessarily 
linearly independent).

We can thus classify eigenfunctions as follows:

\noindent
(i) $\lambda_n=0$ ($H\varphi_0(x)=0$). For this one may impose 
the chirality on~$\varphi_0(x)$ as \begin{equation}
   \gamma_5\varphi_0^\pm(x)=\Gamma_5\varphi_0^\pm(x)
=\pm\varphi_0^\pm(x).
\end{equation}
We denote the number of~$\varphi_0^+(x)$ ($\varphi_0^-(x)$) 
as~$n_+$ ($n_-$).

\noindent
(ii) $\lambda_n\neq0$ and $\lambda_nf(\lambda_n^2)\neq\pm1$. As 
shown above, \begin{equation}
   H\varphi_n(x)=\lambda_n\varphi_n(x),\qquad
   H\widetilde\varphi_n(x)=-\lambda_n\widetilde\varphi_n(x),
\end{equation}
where
\begin{equation}
   \widetilde\varphi_n(x)
   ={1\over\sqrt{1-\lambda_n^2f^2(\lambda_n^2)}
}\Gamma_5\varphi_n(x).
\end{equation}
We have
\begin{equation}
   \Gamma_5\varphi_n(x)
   =\sqrt{1-\lambda_n^2f^2(\lambda_n^2)}
\widetilde\varphi_n(x),\qquad
   \Gamma_5\widetilde\varphi_n(x)
   =\sqrt{1-\lambda_n^2f^2(\lambda_n^2)}\varphi_n(x),
\end{equation}
and
\begin{eqnarray}
   &&\gamma_5\varphi_n(x)=\sqrt{1-\lambda_n^2f^2(\lambda_n^2)}
   \widetilde\varphi_n(x)+\lambda_nf(\lambda_n^2)\varphi_n(x),
\nonumber\\
   &&\gamma_5\widetilde\varphi_n(x)
=\sqrt{1-\lambda_n^2f^2(\lambda_n^2)}
   \varphi_n(x)-\lambda_nf(\lambda_n^2)\widetilde\varphi_n(x).
\end{eqnarray}

\noindent
(iii) $\lambda_nf(\lambda_n^2)=\pm1$, or
\begin{equation}
   H\Psi_\pm(x)=\pm\Lambda\Psi_\pm(x),\qquad\Lambda f(\Lambda^2)
=1. \end{equation} In this case we see \begin{equation}
   \Gamma_5\Psi_\pm(x)=0,
\end{equation}
and
\begin{equation}
   \gamma_5\Psi_\pm(x)=\pm\Lambda f(\Lambda^2)\Psi_\pm(x)
=\pm\Psi_\pm(x).
\end{equation}
We denote the number of~$\Psi_+(x)$ ($\Psi_-(x)$) 
as~$N_+$ ($N_-$). From the relation $\Tr\gamma_5=0$ valid on 
the lattice, one can derive the chirality sum 
rule~\cite{chiu,fujikawa3} 
\begin{equation}
   n_+-n_-+N_+-N_-=0.
\end{equation}
The explicit form of the operator~$H$ is known for 
$f(H^2)=H^{2k}$ with non-negative integers~$k$~\cite{fujikawa}. 
\\
\\
\noindent {\bf Note added}\\
We have emphasized that the free fermion operator 
$1/(1-aD(\gamma_{5}aD)^{2k})$ in (6.1) contains poles at the 
positions of would-be species doublers.
The operator $1/(1-aD(\gamma_{5}aD)^{2k})$ could contain poles 
even in the presence of topologically {\em trivial} gauge 
fields. See, for example,\cite{adams}. If the functional 
measure of  topologically trivial gauge fields which give 
rise to the possible poles is substantial, the domain wall 
representation not only for chiral theory but also for 
vector-like theory would be significantly influenced.

\end{document}